\documentclass[3p,times,sort&compress]{elsarticle}

\usepackage{lineno,graphicx,multirow,hhline,nicefrac}
\usepackage{amsmath,amssymb,color}
\usepackage{epstopdf,pdflscape,float,hyperref,mathrsfs}
\usepackage{nameref,hyperref}
\modulolinenumbers[5]
\journal{}
\newcommand{\RomanNumeralCaps}[1]
{\MakeUppercase{\romannumeral #1}}
\def\cbl{\color{black}}
\def\cb{\color{black}}

\begin{document}

\begin{frontmatter}

\title{Travelling wave analysis of cellular invasion into surrounding tissues.}

\author[qut]{Maud El-Hachem}
\author[qut]{Scott W McCue}
\author[qut]{Matthew J Simpson \corref{cor1}}
\address[qut]{School of Mathematical Sciences, Queensland University of Technology, Brisbane, Australia.}
\cortext[cor1]{Corresponding author: matthew.simpson@qut.edu.au}

\begin{abstract}
Single-species reaction-diffusion equations, such as the Fisher-KPP and Porous-Fisher equations, support travelling wave solutions that are often interpreted as simple mathematical models of biological invasion. Such travelling wave solutions are thought to play a role in various applications including development, wound healing and malignant invasion.  One criticism of these single-species equations is that they do not explicitly describe interactions between the invading population and the surrounding environment.  In this work we study a reaction-diffusion equation that describes malignant invasion which has been used to interpret experimental measurements describing the invasion of malignant melanoma cells into surrounding human skin tissues~\cite{Browning2019}.  This model explicitly describes how the population of cancer cells degrade the surrounding tissues, thereby creating free space into which the cancer cells migrate and proliferate to form an invasion wave of malignant tissue that is coupled to a retreating wave of skin tissue.  We analyse travelling wave solutions of this model using a combination of numerical simulation, phase plane analysis and perturbation techniques.  Our analysis shows that the travelling wave solutions involve a range of very interesting properties that resemble certain well-established features of both the Fisher-KPP and Porous-Fisher equations, as well as a range of novel properties that can be thought of as extensions of these well-studied single-species equations.  Matlab software to implement all calculations is available at \href{https://github.com/ProfMJSimpson/Cellular_Invasion_ElHachem_2021}{GitHub}.
\end{abstract}

\begin{keyword}
Invasion; Cell invasion; Reaction-diffusion; Partial differential equation; Cancer; Phase plane.
\end{keyword}

\end{frontmatter}

\newpage

\section{Introduction}\label{sec:Introduction}
The Fisher-KPP model~\cite{Fisher1937,Kolmogorov1937} is a very simple prototype mathematical model of biological invasion that describes the spatial and temporal evolution of a population where individuals undergo migration by linear diffusion and proliferation via logistic growth. The Porous-Fisher model is an extension of the Fisher-KPP model where the linear diffusion term is generalised to a nonlinear degenerate diffusion term with a power law  diffusivity~\cite{Murray02,Sanchez1994,Witelski1994,Witelski1995,McCue2019}.  Such single species partial differential equation (PDE) models have had a major influence on the study of biological populations, both in cell biology~\cite{Sherratt90,Maini2004,Swanson2003,Sengers2007,Gerlee2012,Jin2020,Largergren2021,Largergren2021b} and in ecology~\cite{Skellam1951,Lewis1994,Holmes1994,Shigesada1995,Steel1998,Kot03,Levin2003}, since these models gives rise to travelling wave solutions that are thought to represent invasion waves~\cite{Canosa1973,Murray02}.  While influential, an obvious limitation of such single species models is that they focus on the properties of the invading population alone, and neglect interactions between the invasive population and the environment, or interactions between the invasive population and other populations of interest.  To overcome this limitation, a number of extended multi-species models that explicitly describe coupling between the invasive population and the environment or other populations of interest have been proposed, for example~\cite{Painter2003,Byrne2003a,Byrne2003b}. \cbl While this work is focused on models of invasion in the context of cancer biology, similar continuum mathematical models are developed and deployed in the ecology literature~\cite{Amor2010,Amor2017,Fort2012,Muller2014}. \cb

\begin{figure}[h]
	\centering
	\includegraphics[width=1\textwidth]{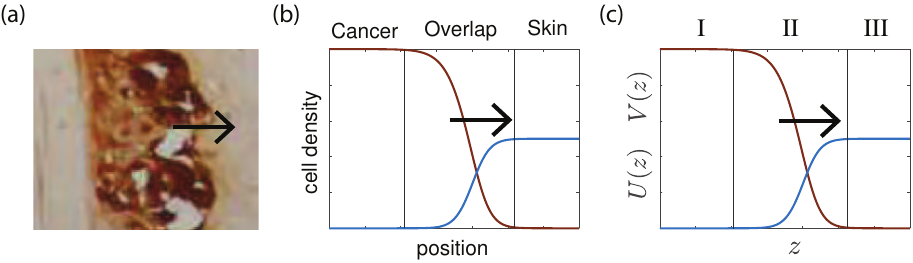}
	\caption{\textbf{Experimental motivation and mathematical model schematic}. (a) Cross section through human skin tissues showing the invasion of melanoma cells (dark brown) into surrounding skin tissue (light brown).  The direction of invasion is shown with the arrow.  This image is reproduced from Haridas~\cite{Haridas2017} with permission. (b) Schematic solution of a one-dimensional PDE model showing the spatial distribution of cell populations during invasion.  To be consistent with the experiments in (a), the density of cancer cells (brown) moves in the direction of the black arrow, and this invasion is associated with the retreat of the density of surrounding skin tissues (blue).  The overlap  region involves a visually-distinct region where both populations are present at the same location.  (c) Shows travelling wave solutions for the cancer cell density $U(z)$, and the skin cell density $V(z)$, on $-\infty<z<\infty$.  In this work we will refer to three different regions of the domain:  (i) region \RomanNumeralCaps 1 is the \textit{invaded} region where $U(z)\rightarrow 1$ and $V(z)\rightarrow 0$ as $z \to -\infty$; (ii) region \RomanNumeralCaps 2 is the \textit{overlap} region where $U(z)>0$ and $V(z)>0$; and, (iii) region \RomanNumeralCaps 3 is the uninvaded region where $U(z)\rightarrow 0$ and $V(z)\rightarrow \mathcal{V}$ as $z \to \infty$.}
	\label{fig:1}
\end{figure}

One area of research where coupled mathematical models of invasion have played an important role is in the study of cancer biology where the invasion of a population of malignant cells is tightly coupled to the degradation of surrounding tissues.  The experimental image in Figure \ref{fig:1}(a) shows a population of highly aggressive melanoma cells growing within and invading into human skin tissues~\cite{Haridas2017,Haridas2018}.  During these experiments, melanoma cells simultaneously migrate and proliferate to form an invading front, and this invasion is tightly coupled to the biochemical breakdown of the surrounding tissues by proteases released by the melanoma cells.  From a mathematical modelling point of view, this kind of process leads to conceptual models like that shown in Figure \ref{fig:1}(b) where we think of a density profile of invading cancer cells that is coupled to the retreat of a population of surrounding skin cells.  In this schematic we identify three regions: \textit{region} \RomanNumeralCaps 1 is the invading region;  \textit{region} \RomanNumeralCaps 2 is the overlap region; and, \textit{region} \RomanNumeralCaps 3 is the uninvaded region.  Throughout this work we refer to the density of invading cells as $\hat{u}$ and the density of surrounding tissues as $\hat{v}$.

One of the first mathematical models of cellular invasion was proposed by Gatenby and Gawlinski~\cite{Gatenby1996}, who present a coupled system of PDEs that described the spatial and temporal development of tumour tissue, normal tissue and excess hydrogen ion concentration.  Numerical exploration and phase plane analysis suggested the formation of a pH gradient across the tumour-host interface, leading to a hypocellular interstitial gap at the tumour-host interface.  This gap was then verified using both \textit{in vitro} and \textit{in vivo} data.  A similar model of cellular migration coupled to the degradation of surrounding tissues in the context of developmental biology was later proposed by Landman and Pettet~\cite{Landman1998}.  In this case the simpler mathematical model was solved exactly to reveal details of the coupling between the invading population and the degradation of surrounding tissues.  Since these first two mathematical models were proposed and analysed in the late 1990s, a range of more detailed mathematical models have since studied to examine different aspects of cellular invasion~\cite{Perumpanani1999,Marchant2000,Smallbone2005,Landman2008,Anderson2008,Astanin2009,Fasano2009,Byrne2010,Tindall2012,Holder2014,Holder2015}.

Despite the fact that mathematical models of cellular invasion into surrounding tissues had been considered for over twenty years, it was not until 2019 that one of these models was first quantitatively calibrated to experimental data.  In 2019, Browning and co-workers~\cite{Browning2019} examined a simplified model based on Gatenby and Gawlinski's earlier, more general modelling framework~\cite{Gatenby1996}.  In this work, Browning took experimental data describing time-series measurements of melanoma invasion into human tissues (Figure 1(a)) and used a Bayesian sequential learning approach to estimate the diffusivity of the melanoma cells, the proliferation rate of the melanoma cells and the rate at which melanoma cells degraded the surrounding skin tissues~\cite{Browning2019}.  This 2019 study was different to many previous mathematical studies of cellular invasion since Browning did not consider any travelling wave solutions or travelling wave analysis.  In the present study we re-visit the model proposed by Browning and explore various travelling wave solutions of that model.  Using a combination of numerical methods to solve the full time-dependent PDE model, phase plane analysis and perturbation methods, we reveal several novel features of the travelling wave solutions of this model.  In particular we unearth many important parallels and differences between the travelling wave solutions of the invasion model and the very well-studied Fisher-KPP model.  There are two aspects of our analysis that are of particular interest.  \cbl First, we show that travelling wave solutions of the invasion model that involves three dimensional phase space can be approximated using the simpler Fisher-KPP phase plane.  Second, we show that Browning's model of invasion leads to travelling wave solutions that are reminiscent of travelling wave solutions of a moving boundary type model~\cite{Ward1997,Byrne1997,Gaffney1999}. \cb

\section{Mathematical model and preliminary simulations}\label{sec:ModelPreliminaryResults}

\subsection{Browning's model of cellular invasion}\label{sec:Model}

In 2019, Browning and co-workers~\cite{Browning2019} proposed the following simple dimensional model to describe the invasion of melanoma cells into surrounding skin tissue,
\begin{align}
\label{eq:GouvdiffUDimensional}
&\dfrac{\partial \hat{u}}{\partial \hat{t}}=\hat{D} \dfrac{\partial}{\partial \hat{x}}\left[\left(1-\dfrac{\hat{v}}{\hat{K}}\right) \dfrac{\partial \hat{u}}{\partial \hat{x}}\right]+\hat{\lambda} \hat{u}\left(1-\dfrac{\hat{u}+\hat{v}}{\hat{K}}\right),& 0 < \hat{x} < \infty,\\
&\dfrac{\partial \hat{v}}{\partial \hat{t}}=-\hat{\gamma} \hat{u} \hat{v},&0 < \hat{x} < \infty, \label{eq:GouvdiffVDimensional}
\end{align}
where $\hat{u}(\hat{x},\hat{t})>0$ is the density of melanoma cells, and $\hat{v}(\hat{x},\hat{t})>0$ is the density of skin cells.  Throughout this work we use a  circumflex to indicate dimensional parameters and variables.

Equation (\ref{eq:GouvdiffUDimensional}) governs the evolution of the cell density, and we see that the melanoma cells move according to a nonlinear diffusion term, with diffusivity  $\hat{D}>0$ $[\mathrm{\mu m}^2 \ \mathrm{h}^{-1}]$.  The nonlinear diffusivity function decreases linearly with the skin density such that the diffusion of melanoma cells vanishes if the skin density reaches the carrying capacity density, $\hat{v}(\hat{x},\hat{t})=\hat{K} > 0 $.  Further, equation (\ref{eq:GouvdiffUDimensional}) specifies that the melanoma cells grow logistically, with rate  $\hat{\lambda}>0$ $[\mathrm{h}^{-1}]$, such that the net proliferation rate is a linearly decreasing function of the total cell density, $\hat{u}(\hat{x},\hat{t})+\hat{v}(\hat{x},\hat{t})$.  Equation (\ref{eq:GouvdiffVDimensional}) governs the evolution of the skin density such that melanoma cells degrade skin cells at a rate governed by $\hat{\gamma} \ge 0$ $[\mathrm{\mu m}^{2}\mathrm{cells}^{-1}\mathrm{h}^{-1}]$.   This two-species PDE model is a simplification of a three-species PDE extension that is fully described in the Supplementary Material document reported by Browning et al.~\cite{Browning2019}.

Since we are interested in travelling wave solutions we pose Equations (\ref{eq:GouvdiffUDimensional})--(\ref{eq:GouvdiffVDimensional}) on $0 < \hat{x} < \infty$, however when we generate numerical solutions we take the usual approach and examine solutions on a truncated domain,  $0 < \hat{x} < \hat{L}$.  For all numerical solutions, we specify no-flux boundary conditions for $\hat{u}(\hat{x},\hat{t})$, so that $\partial \hat{u}(0,\hat{t}) /\partial \hat{x} =  \partial \hat{u}(\hat{L},\hat{t}) /\partial \hat{x} = 0$.
Note that since Equation (\ref{eq:GouvdiffVDimensional}) does not involve any spatial derivatives, we do not need to specify any boundary conditions for $\hat{v}(\hat{x},\hat{t})$.   The choice of $\hat{L}$ is unimportant provided that it is chosen to be sufficiently large~\cite{ElHachem2021}.  Matlab software on  \href{https://github.com/ProfMJSimpson/Cellular_Invasion_ElHachem_2021}{GitHub} can be used to explore different choices of $\hat{L}$ for all problems that we consider, and full details of the numerical algorithms are outlined in the Appendix.

The mathematical model is nondimensionalised by introducing $u=\hat{u} /\hat{K}$,  $v=\hat{v} / \hat{K}$, $x = \hat{x}\sqrt{\hat{\lambda}/\hat{D}}$, $t = \hat{\lambda} \hat{t}$, and $\gamma = \hat{K}\hat{\gamma}/\hat{\lambda}$, which gives
\begin{align}
\label{eq:GouvdiffUNonDimensional}
&\dfrac{\partial u}{\partial t}= \dfrac{\partial}{\partial x}\left[(1-v) \dfrac{\partial u}{\partial x}\right] + u(1-u-v), &0 < x < \infty\\
&\dfrac{\partial v}{\partial t}=-\gamma uv, &0 < x < \infty, \label{eq:GouvdiffVNonDimensional}
\end{align}
which means that the nondimensional PDE model requires the specification of just one model parameter, $\gamma \ge 0$.

For the first part of the study, we specify initial conditions such that the initial distribution of skin density is a constant, and the initial distribution of melanoma cells has compact support,
\begin{align}
	u(x,0) &= \alpha \left[1 - H(\beta) \right], \label{eq:ICUCompactSupport}\\
	v(x,0) &= \mathcal{V},	 \label{eq:ICVCompactSupport}
\end{align}
where $H(x)$ is the usual Heaviside function and $\beta > 0$ is a constant so that we have $u(x,0)=\alpha$ for $x < \beta$ and $u(x,0)=0$ for $x > \beta$.  The initial density of skin, $0 \le \mathcal{V}  \le 1$, is set to be a constant that does not depend upon position.  In summary, the non-dimensional invasion model involves one model parameter, $\gamma \ge 0$, and when we specify the initial conditions we introduce another parameter, $0 \le \mathcal{V} \le 1$, that we will study.  In general, we interpret $\gamma$ as the rate at which cancer cells degrade skin tissues, and $\mathcal{V}$ is the density of skin tissues in the far field ahead of the invading front.  Most of our analysis will be concerned with how varying $\gamma$ and $\mathcal{V}$ influence the shape and speed of the resulting travelling wave solutions. Setting the upper value of $\mathcal{V}=1$ implies that the maximum density of skin cells is the same as the maximum density of cancer cells. This assumption may be biologically realistic as skin cells and cancer cells are often similar in shape and size~\cite{Haridas2017}.

\cbl The first part of this work focuses on travelling wave solutions of the invasion model that arise from initial conditions given by Equations (\ref{eq:ICUCompactSupport})--(\ref{eq:ICVCompactSupport}) since this form of initial condition with compact support is biologically-relevant~\cite{Haridas2017,Haridas2018}.  Once we have characterised this first family of travelling wave solutions, we will then study another family of travelling wave solutions where $u(x,0)$ decays to zero as $x \to \infty$.  While this second family of travelling wave solutions are mathematically interesting, the fact that these travelling wave solutions that arise from exponentially decaying initial conditions means that this second set of travelling waves are more difficult to interpret biologically since this kind of special initial condition is not relevant in practice. \cb  To study this second family of travelling wave solutions we work with an initial condition given by
\begin{align}
\label{eq:ICUNonCompactSupport}
u(x,0) &=
	\begin{cases}
	\alpha, &x<\beta, \\
	\alpha \, \textrm{exp}[-a(x-\beta)], &x\ge\beta,
	\end{cases}\\
v(x,0) &= \mathcal{V},	 \label{eq:ICVNonCompactSupport}
\end{align}
where $a > 0$ is a constant that we vary so that we can understand how travelling wave solutions of the invasion model depend upon the initial spatial decay of $u$.

One of the main themes of our work is to highlight surprising similarities and differences between the invasion model   (\ref{eq:GouvdiffUNonDimensional})--(\ref{eq:GouvdiffVNonDimensional}) and some well-studied single-species models.  For example, setting $\mathcal{V} = 0$ reduces the invasion model to the well-known Fisher-KPP model~\cite{Fisher1937,Kolmogorov1937,Murray02},
\begin{equation}
\label{eq:FisherKPPNonDimensional}
\dfrac{\partial u}{\partial t}= \dfrac{\partial^2 u}{\partial x^2} + u(1-u).
\end{equation}
As we pointed out in Section \ref{sec:Introduction}, the Fisher-KPP model is a simplified mathematical model of biological invasion that ignores any interaction between the invading population and the local environment.  In this work, we study travelling wave solutions of Equations (\ref{eq:GouvdiffUNonDimensional})--(\ref{eq:GouvdiffVNonDimensional}) and we find there are important similarities and differences with travelling wave solutions of the Fisher-KPP model, so it is useful to explicitly note the connection between these two mathematical models at the outset.  There are also interesting, but perhaps less obvious, connections between the invasion model and the Porous-Fisher model~\cite{Murray02,Sanchez1994,Witelski1994,Witelski1995,McCue2019},
\begin{equation}
\label{eq:PorousFisherNonDimensional}
\dfrac{\partial u}{\partial t}= \dfrac{\partial}{\partial x}\left[u \dfrac{\partial u}{\partial x} \right] + u(1-u).
\end{equation}
We will now begin to explore these relationships.

\subsection{Time-dependant solutions}\label{sec:TimeDependantSolution}

We begin our study of the invasion model by computationally exploring various travelling wave solutions of (\ref{eq:GouvdiffUNonDimensional})--(\ref{eq:GouvdiffVNonDimensional}) to develop an intuitive understanding of how their shape and speed depend upon the choice of $\gamma$ and $\mathcal{V}$.  Full details of the numerical method used to solve the PDE model is given in the Appendix, and MATLAB software to study time-dependent solutions of  (\ref{eq:GouvdiffUNonDimensional})--(\ref{eq:GouvdiffVNonDimensional}) is available on \href{https://github.com/ProfMJSimpson/Cellular_Invasion_ElHachem_2021}{GitHub}.

\begin{landscape}
	\begin{figure}[h!]
		\centering
		\includegraphics[height=0.65\textheight]{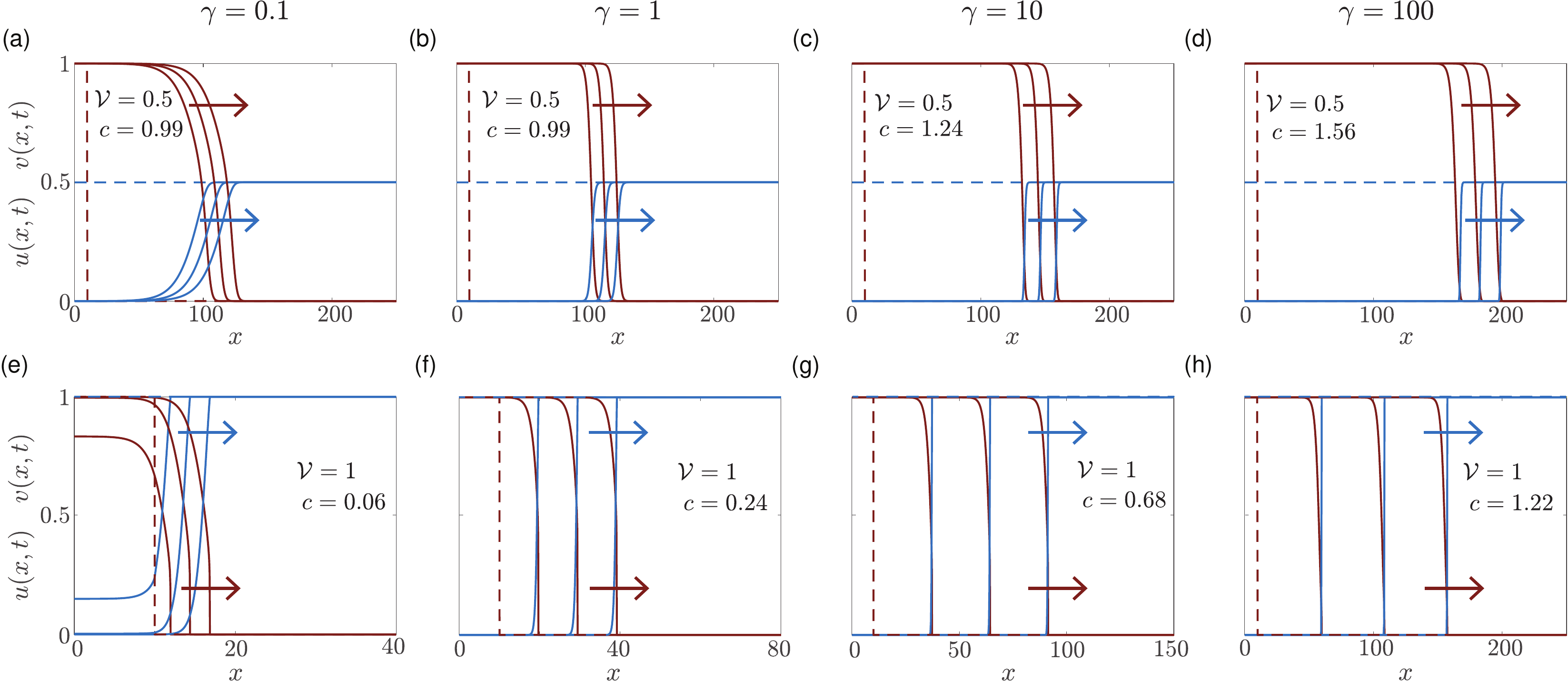}
		\caption{\textbf{Time-dependant solutions of Equations (\ref{eq:GouvdiffUNonDimensional})--(\ref{eq:GouvdiffVNonDimensional}) for different $\gamma$ and $\mathcal{V}$.} Results in (a)--(d) correspond to $\mathcal{V}=0.5$, while results in (e)--(h) correspond to $\mathcal{V}=1$.  Solutions in (a) and (e), (b) and (f), (c) and (g), and (d) and (h) correspond to $\gamma=0.1, 1, 10$ and $100$, respectively.   Density profiles for $u(x,t)$ are shown in brown, and profiles of $v(x,t)$ are shown in blue. Results in (a)--(d) are shown at $t=0, 100, 110, 120$ and the invasion front moves in the positive $x$-direction, whereas results in (e)--(h) are shown at $t=0, 40, 80, 120$.  For each solution we show the initial condition in dashed lines, and the subsequent solutions for $t > 0$ are shown in solid lines. All PDE solutions are obtained using $\Delta x = \Delta t = 1 \times 10^{-3}$. }
		\label{fig:2}
	\end{figure}
\end{landscape}

In the first instance we focus on travelling wave solutions with initial condition given by Equations (\ref{eq:ICUCompactSupport})--(\ref{eq:ICVCompactSupport}) so that the initial cancer density profile has compact support.  Time-dependent solutions are shown in Figure \ref{fig:2}(a)--(d) for $\mathcal{V}=0.5$ for a range of $\gamma$.  Additional time-dependent solutions are shown in Figure \ref{fig:2}(e)--(h) for $\mathcal{V}=1$ for a range of $\gamma$.  In all cases we see that the time-dependent solutions of Equations (\ref{eq:GouvdiffUNonDimensional})--(\ref{eq:GouvdiffVNonDimensional}) evolve to constant-speed travelling wave solutions, and in each subfigure we give a numerical estimate of the travelling wave speed, $c$.

Results in Figure \ref{fig:2}(a)--(d) highlight several interesting properties of these travelling wave solutions, especially when we compare them with the well-known travelling wave solutions of the Fisher-KPP model (\ref{eq:FisherKPPNonDimensional}).  Each travelling wave solution in Figure \ref{fig:2}(a)--(d) leads to profiles for $u(x,t)$ that \cbl are smooth since they are differentiable everywhere on the domain, and they do not have compact support since they decay to zero in the far field, $u(x,t) \to 0^+$ as $x \rightarrow \infty$.  In this regard, the shape of these travelling wave solutions is similar to those for the Fisher-KPP model~\cite{Murray02,Canosa1973}. \cb However, in the invasion model we see that the speed of the travelling wave depends upon the decay rate, $\gamma$, in a rather unexpected way.  First, results in Figure \ref{fig:2}(a)--(b) suggest that for sufficiently small $\gamma < \gamma_{\textrm{c}}$, the speed of the travelling wave is independent of $\gamma$.  Second, results in Figure \ref{fig:2}(c)--(d) indicate that for sufficiently large $\gamma > \gamma_{\textrm{c}}$, the travelling wave speed increases with $\gamma$.  It is of interest to note that all values of $\gamma$ considered in Figure \ref{fig:2}(a)--(d) lead to travelling wave solutions with $c < 2$.  This is very different to travelling wave solutions of the dimensionless Fisher-KPP model that evolve from initial conditions with compact support since these travelling waves always have $c = 2$~\cite{Murray02,Canosa1973}.

Results in Figure \ref{fig:2}(e)--(h) for $\mathcal{V}=1$ show that the invasion model leads to non-smooth travelling wave solutions that have compact support. \cbl These solutions are non-smooth since they contain a well defined front and $u(x,t)$ is not differentiable at the location of the front, $x=X$ (sometimes called the contact point). These solutions have compact support since $u(x,t) > 0$ for $x < X$, and $u(x,t)=0$ for $x \ge X$. \cb   From this point of view, the shape of the travelling wave solutions in Figure \ref{fig:2}(e)--(h) is reminiscent of the shape of travelling wave solutions of the Porous-Fisher model (\ref{eq:PorousFisherNonDimensional}). Comparing the speeds of the travelling wave solutions in Figure \ref{fig:2}(e)--(h) indicates that $c$ increases with $\gamma$ but, unlike the results in Figure \ref{fig:2}(a)--(d), we do not see any evidence that the wave speed is independent of $\gamma$ for any of the values considered.  As with Figure \ref{fig:2}(a)--(d), all values of $\gamma$ explored in Figure \ref{fig:2}(e)--(h) lead to travelling wave solutions with $c < 2$, which, again, is very different to travelling wave solutions of the dimensionless Fisher-KPP equation (\ref{eq:FisherKPPNonDimensional}).

\begin{figure}[h!]
	\centering
	\includegraphics[width=1\textwidth]{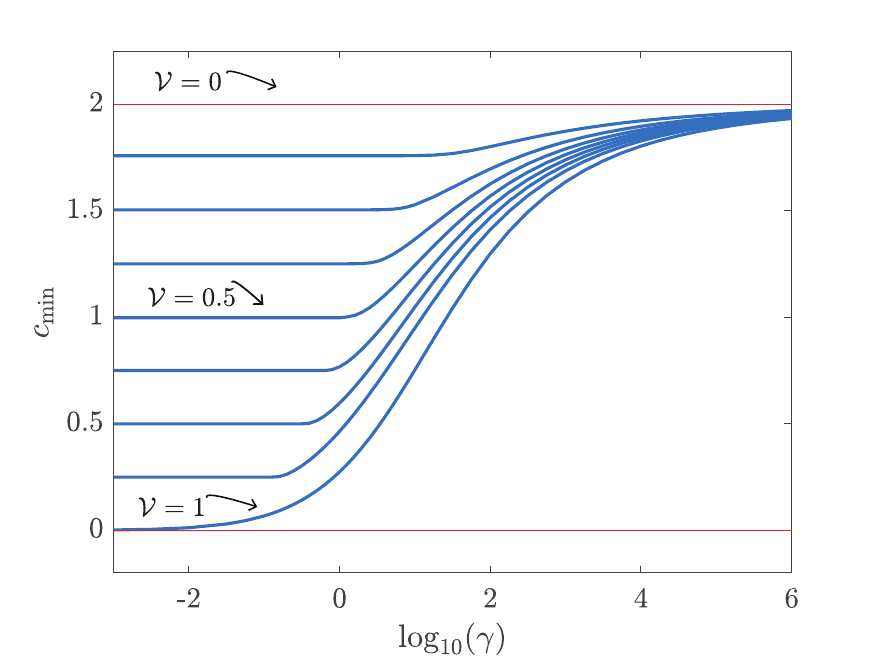}
	\caption{\textbf{Relationship between the minimum travelling wave speed $c_{\textrm{min}}$,  $\gamma$ and $\mathcal{V}$.} Numerical estimates of $c_{\textrm{min}}$ are obtained by solving Equations (\ref{eq:GouvdiffUNonDimensional})--(\ref{eq:GouvdiffVNonDimensional}) with initial conditions (\ref{eq:ICUCompactSupport})--(\ref{eq:ICVCompactSupport}), with $\alpha = 1$ and $\beta = 10$. Numerical solutions of the PDE model are obtained with $\Delta x = \Delta t= 1 \times 10^{-2}$, for $1\times10^{-3}\le\gamma\le1\times10^{6}$ and $\mathcal{V} = 1/8, 2/8, 3/8, 4/8, 5/8, 6/8, 7/8$ and $1$.}
	\label{fig:3}
\end{figure}

In summary, preliminary numerical explorations reveal that we obtain smooth travelling wave solutions of the invasion model for $\mathcal{V} < 1$ whereas we obtain non-smooth travelling waves with compact support when $\mathcal{V}=1$.  While the former are similar in shape to the travelling wave solutions of the Fisher-KPP model and the latter are similar in shape to the travelling wave solutions of the Porous-Fisher model, the speed of the travelling waves of the invasion model are very different to the well-known bounds on travelling wave solutions of the Fisher-KPP and Porous-Fisher models. \cbl  The relationship between $c$, $\gamma$ and $\mathcal{V}$ is summarised in Figure \ref{fig:3} where we estimate the long time travelling wave speed $c$ from a large number of simulations where $u(x,0)$ has compact support, given by Equations (\ref{eq:ICUCompactSupport})--(\ref{eq:ICVCompactSupport}.  As we will establish later in Section \ref{sec:CGreaterThanCmin}, this means that these estimates of $c$ correspond to the minimum wave speed, $c_{\textrm{min}}$.  Remembering that the Fisher-KPP model has a very simple minimum wave speed,  $c_{\textrm{min}} = 2$~\cite{Murray02,Canosa1973}, here we see that the minimum wave speed for the invasion model is a relatively complicated function of $\mathcal{V}$ and $\gamma$ and we will devote much of this work to understanding this relationship in different limiting cases. \cb  As noted in Section \ref{sec:Model}, setting $\mathcal{V}=0$ means that the invasion model simplifies to the Fisher-KPP model and so we obtain $c=c_{\textrm{min}}=2$ regardless of $\gamma$.  Interestingly, for all choices of $\gamma$ and $\mathcal{V} > 0$ we observe travelling wave solutions with a minimum wave speed that is less than the minimum travelling wave speed for the Fisher-KPP model, and we will return to this point later.  For intermediate values of $0 < \mathcal{V} < 1$ we see that $c_{\textrm{min}}$ is a decreasing function of $\mathcal{V}$, but independent of $\gamma$ provided that $\gamma$ is sufficiently small.  In contrast for intermediate values of $0 < \mathcal{V} < 1$ we see that $c_{\textrm{min}}$ increases with $\gamma$, and approaches  $c_{\textrm{min}}= 2$ for large $\gamma$.  Finally, for $\mathcal{V}=1$, there appears to be no positive value of $\gamma$ where the wave speed is independent of $\gamma$, meaning that we have with $c_{\textrm{min}} \to 0^+$ as $\gamma \to 0^+$, and $c_{\textrm{min}} \to 2^-$ as $\gamma \to \infty$. The key features of the travelling wave solutions of the invasion model, and their relationship to the well-studied travelling wave solutions of the Fisher-KPP and Porous-Fisher models are summarised in Table \ref{tab:ComparisonFisherKPPInvasionModel}.

\begin{table}[h!]
	
\caption{Comparing key features of travelling wave solutions of the invasion model with travelling wave solutions of the Fisher-KPP model and the Porous-Fisher model.}	
\scalebox{1.0}{
	\centering
	\begin{tabular}{|c|c|c|c|c| }
		\hline
		\multirow{2}{*}{Solution features} & \multirow{2}{*}{Fisher-KPP (\ref{eq:FisherKPPNonDimensional})} & \multirow{2}{*}{Porous-Fisher (\ref{eq:PorousFisherNonDimensional})} & \multicolumn{2}{|c|}{Invasion model}\\
		\cline{4-5}
		& & & $\mathcal{V}<1$ & $\mathcal{V}=1$ \\
		\hhline{|=|=|=|=|=|}
		\multirow{2}{*}{Travelling wave shape} & Smooth & Sharp-fronted & Smooth & Sharp-fronted  \\
		\cline{2-5}
	 &No compact support & Compact support & No compact support & Compact support \\
		\hline
		\multirow{2}{*}{Minimum wave speed}& \multirow{2}{*}{$c_\textrm{min} = 2$} & \multirow{2}{*}{$c_\textrm{min} = \dfrac{1}{\sqrt{2}}$} & $c_\textrm{min} = 2(1-\mathcal{V})$, \, $\gamma < \gamma_{\textrm{c}}$ &  $\displaystyle{\lim_{\gamma \to 0^+}c_\textrm{min} = 0^+}$ \\
		\cline{4-5}
	&	& & $\displaystyle{\lim_{\gamma \to \infty}c_\textrm{min} = 2^-}$ &  $\displaystyle{\lim_{\gamma \to \infty}c_\textrm{min} = 2^-}$ \\
		\hline
	\end{tabular}
\label{tab:ComparisonFisherKPPInvasionModel}
	}
\end{table}

\newpage

\section{Travelling wave analysis}\label{sec:TravellingWaveSol}
We now attempt to formalise these preliminary numerical results by analysing travelling wave solutions of Equations (\ref{eq:GouvdiffUNonDimensional})--(\ref{eq:GouvdiffVNonDimensional}) in the phase-space.

\subsection{Preamble}\label{sec:Preamble}

To study the travelling wave solutions of Equations (\ref{eq:GouvdiffUNonDimensional})--(\ref{eq:GouvdiffVNonDimensional}) we seek solutions in the form $u(x, t)= U(z)$ and $V(x, t)=V(z)$, where $z$ is the usual travelling wave variable, $z=x-ct$.  Re-writing the governing equations in terms of the travelling wave coordinate gives
\begin{align}
\dfrac{\mathrm{d}}{\mathrm{d}z}\left[(1-V)\dfrac{\mathrm{d}U}{\mathrm{d}z}\right] + c\dfrac{\mathrm{d}U}{\mathrm{d}z} + U(1-U-V) &= 0,&-\infty < z < \infty, \label{eq:ODEUz}\\
c\dfrac{\mathrm{d} V}{\mathrm{d} z} -  \gamma U V &= 0,&-\infty < z < \infty, \label{eq:ODEVz}
\end{align}
with boundary conditions $U(z) \rightarrow 1$ and $V(z) \rightarrow 0$ as $z\rightarrow -\infty$, and $U(z) \rightarrow 0$ and $V(z) \rightarrow \mathcal{V}$ as $z\rightarrow \infty$. At this point it is worthwhile to observe that if the solution $U(z)$ is known, we can solve Equation (\ref{eq:ODEVz}) to give,
\begin{equation}
V(z) = \mathcal{V} \exp{\left(\dfrac{\gamma}{c}\int_{z}^{\infty}  U(\xi) \ \mathrm{d} \xi \right)}. \label{eq:SolutionVzIntegralU}
\end{equation}

To study this boundary value problem in  phase space we re-write Equations (\ref{eq:ODEUz})--(\ref{eq:ODEVz}) as a first order system
\begin{align}
\label{eq:ODEdUdz}
\dfrac{\mathrm{d} U}{\mathrm{d} z} &= W,\\
\dfrac{\mathrm{d} V}{\mathrm{d} z} &= \dfrac{\gamma U V}{c}, \label{eq:ODEdVdz}\\
\dfrac{\mathrm{d} W}{\mathrm{d} z} &= \dfrac{1}{(1-V)}\left[\dfrac{\gamma U V W}{c} - cW - U(1-U-V)\right], \label{eq:ODEdWdz}
\end{align}
with boundary conditions $U \rightarrow 1, V \rightarrow 0$ and $W \rightarrow 0$ as $z\rightarrow -\infty$, and $U \rightarrow 0, V \rightarrow  \mathcal{V}$ and $W \rightarrow 0$ as $z\rightarrow \infty$.  There are two equilibrium points of this dynamical system: (i) $(\bar{U}, \bar{V}, \bar{W}) = (1,0,0)$ corresponding to $z \to -\infty$; and, (ii) $(\bar{U}, \bar{V}, \bar{W}) = (0,\mathcal{V},0)$ for $\mathcal{V} < 1$, corresponding to $z \to \infty$.   Before we proceed it is useful to remark that the dynamical system is singular when $\mathcal{V}=1$, whereas there is no such singularity for $\mathcal{V}< 1$.  Therefore, just like we did in Section \ref{sec:ModelPreliminaryResults}, we will treat these two cases separately.

\subsection{$\mathcal{V} < 1$} \label{sec:PhasePlaneVsmallerThan1}
Setting $\mathcal{V} < 1$ leads to smooth travelling wave solutions that we will analyse in phase space. \cbl  Since the travelling wave solutions are smooth the phase plane is non-singular.  We refer to this as a traditional phase space since we do not have to consider any nonsingularities, and we proceed by analysing Equations (\ref{eq:ODEdUdz})--(\ref{eq:ODEdWdz}) directly. \cb  Eigenvalues of the linearised dynamical system about $(\bar{U}, \bar{V}, \bar{W}) = (1,0,0)$ are given by the  roots of $\lambda^3-(\gamma-c^2)\lambda^2/c-(1+\gamma)\lambda+\gamma/c = 0$, so that we have  $\lambda_1 =  \gamma/c$ and $\lambda_{2,3} = (-c \pm \sqrt{c^2+4})/2)$. These eigenvalues are all real with $\lambda_1 > 0$, $\lambda_2 > 0$ and $\lambda_3 < 0$, which means that the equilibrium point $(\bar{U}, \bar{V}, \bar{W}) = (1,0,0)$ is a three-dimensional saddle for all values of $c$ and $\gamma$ \cite{Wiggins2003}.   Note that the analogous phase plane analysis for travelling wave solutions of the Fisher-KPP model (\ref{eq:FisherKPPNonDimensional}) involves an equilibrium point corresponding to the invaded region that is a two-dimensional saddle for all $c$~\cite{Murray02}.

Eigenvalues of the linearised dynamical system about $(\bar{U}, \bar{V}, \bar{W}) = (0,\mathcal{V},0)$ are given by the roots of $\lambda^3+c\lambda^2/(\mathcal{V}-1)+\lambda = 0$, so that we have  $\lambda_1 = 0$ and $\lambda_{2,3}= (-c \pm \sqrt{c^2-4(1-\mathcal{V})^2})/[2(1-\mathcal{V})]$. In this case $\lambda_2$ and $\lambda_3$ are real negative numbers when $c \ge 2(1-\mathcal{V})$, whereas they are complex conjugates when $c < 2(1-\mathcal{V})$.  This means that the equilibrium point associated with the uninvaded region is a non-hyperbolic stable node when $c \ge 2(1-\mathcal{V})$ and a non-hyperbolic stable spiral when $c < 2(1-\mathcal{V})$.  Since we are interested in travelling waves with $U(z) > 0$, this condition defines a minimum wave speed, $c_\textrm{min} =  2(1-\mathcal{V})$.  Again, note that the analogous phase plane analysis of the Fisher-KPP model also involves the equilibrium point corresponding to the uninvaded region bifurcating from a stable node to a stable spiral, and this defines an analogous minimum wave speed, $c_{\textrm{min}}=2$~\cite{Murray02}.

The phase space analysis leading to a minimum wave speed condition, $c_\textrm{min} = 2(1-\mathcal{V})$ is consistent with our numerical results in Figure \ref{fig:3} computed with initial conditions with compact support.  In particular, for sufficiently small $\gamma < \gamma_{\textrm{c}}$ we have travelling wave solutions that move with the minimum wave speed, $c_\textrm{min} = 2(1-\mathcal{V})$, and this speed is independent of $\gamma$.  For large $\gamma > \gamma_{\textrm{c}}$ our numerical results lead to travelling wave solutions with $c > c_\textrm{min} = 2(1-\mathcal{V})$, which is again consistent with the phase space analysis.  Unfortunately, this analysis provides no insight into the behaviour of the wave speed for sufficiently large $\gamma$, nor the critical value, $\gamma_{\textrm{c}}$, where the wave speed dependence upon $\gamma$ changes. \cbl Alternatively, simply
set $\gamma = 0$ in Equations (\ref{eq:GouvdiffUNonDimensional})--(\ref{eq:GouvdiffVNonDimensional}) uncouples the system.  Seeking travelling wave solutions of this uncoupled system leads to Fisher-KPP-like phase plane analysis, giving   $c_\textrm{min}=2(1-\mathcal{V})$ which corroborates our previous result.   Unfortunately, this simpler approach provides no insight when $\gamma > 0$.\cb

\cbl Biologically, setting $\mathcal{V} < 1$ corresponds to the situation where the density of the skin tissue ahead of the invading front of is lower than the carrying capacity of the invading population. Intuitively, we might anticipate that the speed of the invading front would decrease with $\mathcal{V}$ and increase with $\gamma$.  While the first of these intuitive expectations is consistent with our analysis and numerical observations, the second point highlights the value of our mathematical analysis and numerical explorations since our finding that the minimum wave speed is independent of $\gamma$, for sufficiently small $\gamma < \gamma_{\textrm{c}}$, is not at all obvious.  The implication of this finding is that interventions seeking to reduce the decay rate to zero would not stop the invasion since $c_{\textrm{min}} > 0$ when $\gamma=0$.\cb

\subsection{$\mathcal{V} = 1$}\label{sec:PhaseplaneVequal1}
Setting $\mathcal{V} = 1$ leads to nonsmooth travelling wave solutions and a singularity in Equations (\ref{eq:ODEdUdz})--(\ref{eq:ODEdWdz}).  We proceed by defining a new independent variable $\zeta$ , $(1-V)\mathrm{d}(\cdot)/\mathrm{d} z= \mathrm{d} (\cdot)/ \mathrm{d} \zeta$~\cite{Murray02}, so that the desingularised dynamical system is given by
\begin{align} \label{eq:ODEdUdzeta}
\dfrac{\mathrm{d}U}{\mathrm{d}\zeta} &= W(1-V),\\
\dfrac{\mathrm{d}V}{\mathrm{d}\zeta} &= \left(\dfrac{\gamma U V }{c}\right)(1-V),\label{eq:ODEdVdzeta}\\
\dfrac{\mathrm{d}W}{\mathrm{d}\zeta} &=\left(\dfrac{\gamma UV - c^2}{c}\right)W - U(1-U-V),\label{eq:ODEdWdzeta}
\end{align}
with boundary conditions $U \rightarrow 1, W \rightarrow 0$ and $V \rightarrow 0$ as $\zeta \rightarrow -\infty$, and $U \rightarrow 0, W \rightarrow 0$ and $V \rightarrow \  \mathcal{V}$ as $\zeta \rightarrow \infty $.  Similar to before, given $U(\zeta)$, the solution for $V(\zeta)$ can be written as
\begin{equation}
V(\zeta) = \dfrac{1}{\displaystyle{A\exp{\left(-\dfrac{\gamma}{c}\int_{\zeta}^{\infty}  U(\eta) \ \mathrm{d}\eta\right) }}+1}, \label{eq:Vzetaexp}
\end{equation}
where $A$ is an integration constant.

Eigenvalues of the linearised dynamical system around $(\bar{U}, \bar{V}, \bar{W}) = (1,0,0)$ are roots of $\lambda^3-(\gamma-c^2)\lambda^2/c-(1+\gamma)\lambda+\gamma/c = 0$, leading to $\lambda_1 =  \gamma/c$, $\lambda_{2,3} = (-c \pm \sqrt{c^2+4})/2)$.  This means that $\lambda_2$ and $\lambda_3$ are real numbers with opposite sign, and so $(\bar{U}, \bar{V}, \bar{W}) = (1,0,0)$ is a saddle.  The eigenvalues of the linearised dynamical system around $(\bar{U}, \bar{V}, \bar{W}) = (0,1,0)$ are roots of $c\lambda^2+\lambda^3=0$, leading to $\lambda_1 = -c$,  $\lambda_{2} = \lambda_{3} = 0$, which means that the uninvaded equilibrium point is always a degenerate stable node.  Unlike the previous case where $\mathcal{V} < 1$, here there is no restriction on a minimum wave speed to ensure $U(\zeta) > 0$.  One way of interpreting this result is that $c_\textrm{min}=0$ which is consistent with the numerical results in Figure \ref{fig:3} where we have $c > c_\textrm{min}=0$ when $\mathcal{V} = 1$.

\cbl Biologically, setting $\mathcal{V} = 1$ corresponds to the situation where the density of the skin tissue ahead of the invading front is identical to the carrying capacity of the invading population. In this situation we see that $c_{\textrm{min}} \to 0^+$ as $\gamma \to 0^+$, such that $\gamma_{\textrm{c}}=0$ and there is no interval of $\gamma$ for which the minimum wave speed is independent of $\gamma$. The implication of this finding is that interventions seeking to reduce the decay rate to $\gamma=0$ would eventually stop the invasion since $c_{\textrm{min}} \to 0^+$ as $\gamma \to 0^+$. \cb

\section{Limiting cases}\label{sec:SpecialLimit}
As we pointed out in Section \ref{sec:Model}, we are primarily interested in understanding how travelling wave solutions of the invasion model (\ref{eq:GouvdiffUNonDimensional})--(\ref{eq:GouvdiffVNonDimensional}) depend upon choices of $\gamma$ and $\mathcal{V}$. We will start by considering limits of fast and slow decay, $\gamma \gg 1$ and $\gamma \ll 1$, respectively, and consider differences between $\mathcal{V} = 1$ and $\mathcal{V} < 1$ as appropriate.

\subsection{Fast decay: $\gamma \gg 1$}\label{sec:GammaInfinity}
Preliminary numerical results in Figure \ref{fig:2} indicate that the width of the overlap region (region \RomanNumeralCaps 2) decreases with $\gamma$.  This trend is evident in Figure \ref{fig:2}(a)--(d) for  $\mathcal{V} < 1$ as well as in Figure \ref{fig:2}(e)--(h) where $\mathcal{V}=1$.  One way to interpret this observation is that the overlap between the $U(z)$ and $V(z)$ profiles becomes negligible as $\gamma$ increases.

Given our previous discussion in Section \ref{sec:Model} where we observed that setting $\mathcal{V} = 0$ means that the evolution equation for $u(x,t)$ simplifies to the Fisher-KPP model (\ref{eq:FisherKPPNonDimensional}), we anticipate that the solution of the dynamical system associated with travelling wave solutions of the invasion model for $\gamma \gg 1$ can be approximated by the solution of the dynamical system associated with travelling wave solutions of the Fisher-KPP model,
\begin{equation}
\dfrac{\mathrm{d}^2U}{\mathrm{d}z^2} + c\dfrac{\mathrm{d}U}{\mathrm{d}z} + U(1-U) = 0, \label{eq:ODEFKPP}
\end{equation}
with boundary conditions $U \rightarrow 1$ as $z \rightarrow -\infty$, and $U \rightarrow 0$ as $z \rightarrow \infty$.   In the usual way, this second-order boundary value problem can be re-written in terms of a first order system
\begin{align}
\label{eq:ODEdUdzFKPP}
\dfrac{\mathrm{d} U}{\mathrm{d} z} &= W,\\
\dfrac{\mathrm{d} W}{\mathrm{d} z} &= -cW - U(1-U), \label{eq:ODEdWdzFKPP}
\end{align}
with boundary conditions $U \rightarrow 1$ and $W \rightarrow 0$ as $z\rightarrow -\infty$, and $U \rightarrow 0$ and $W \rightarrow 0$ as $z\rightarrow -\infty$.  There are two equilibrium points: (i) $(\bar{U},  \bar{W}) = (1, 0)$ that is associated with the invaded region;  and, (ii) $(\bar{U}, \bar{W}) = (0,0)$ that is associated with the uninvaded region.

\begin{landscape}
	\begin{figure}[h!]
		\centering
		\includegraphics[width=1\linewidth]{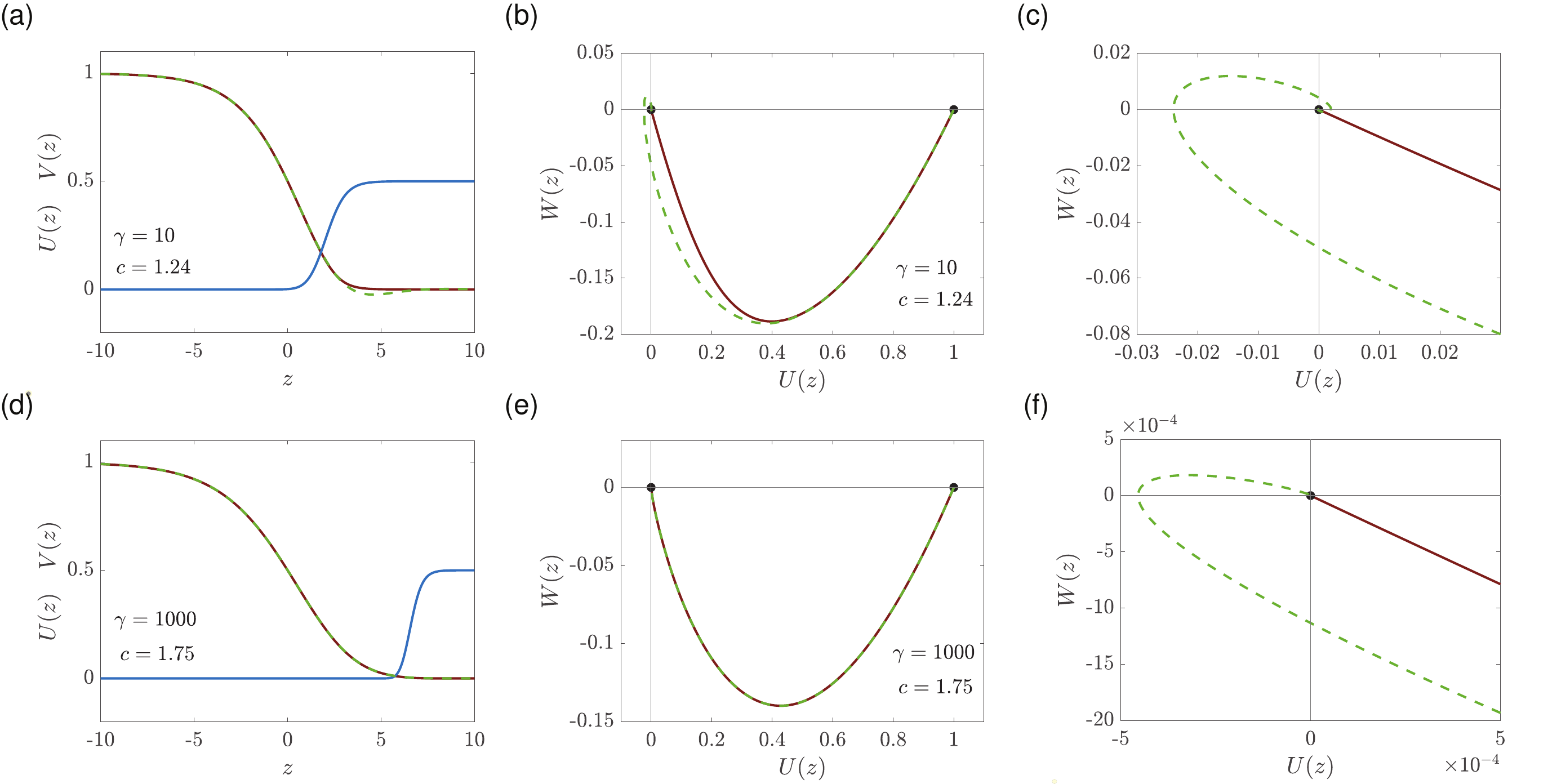}
		\caption{\textbf{Travelling wave solutions of the invasion model with $\gamma \gg 1$ and $\mathcal{V} < 1$.}  (a) and (d) show late-time numerical solutions of Equations (\ref{eq:GouvdiffUNonDimensional})--(\ref{eq:GouvdiffVNonDimensional}) with $\gamma=10$ and $\gamma=1000$, respectively.  The solution for $U(z)$ is shown in solid brown and the solution for $V(z)$ is shown in solid blue.  The approximate solution for $U(z)$ obtained from the Fisher-KPP phase plane is shown in dashed green.  (b) and (c) show a projection of the $(U,V,W)$ phase space for the invasion model projected onto the $(U,W)$ plane together with the projection of the three-dimensional heteroclinic orbit in solid brown.  The trajectory from Fisher-KPP phase plane is shown in dashed green.  (c) and (f) show magnified regions near the origin in (b) and (e), respectively.}
		\label{fig:4}
	\end{figure}	
\end{landscape}

\begin{landscape}
	\begin{figure}[h!]
		\centering
		\includegraphics[width=1\linewidth]{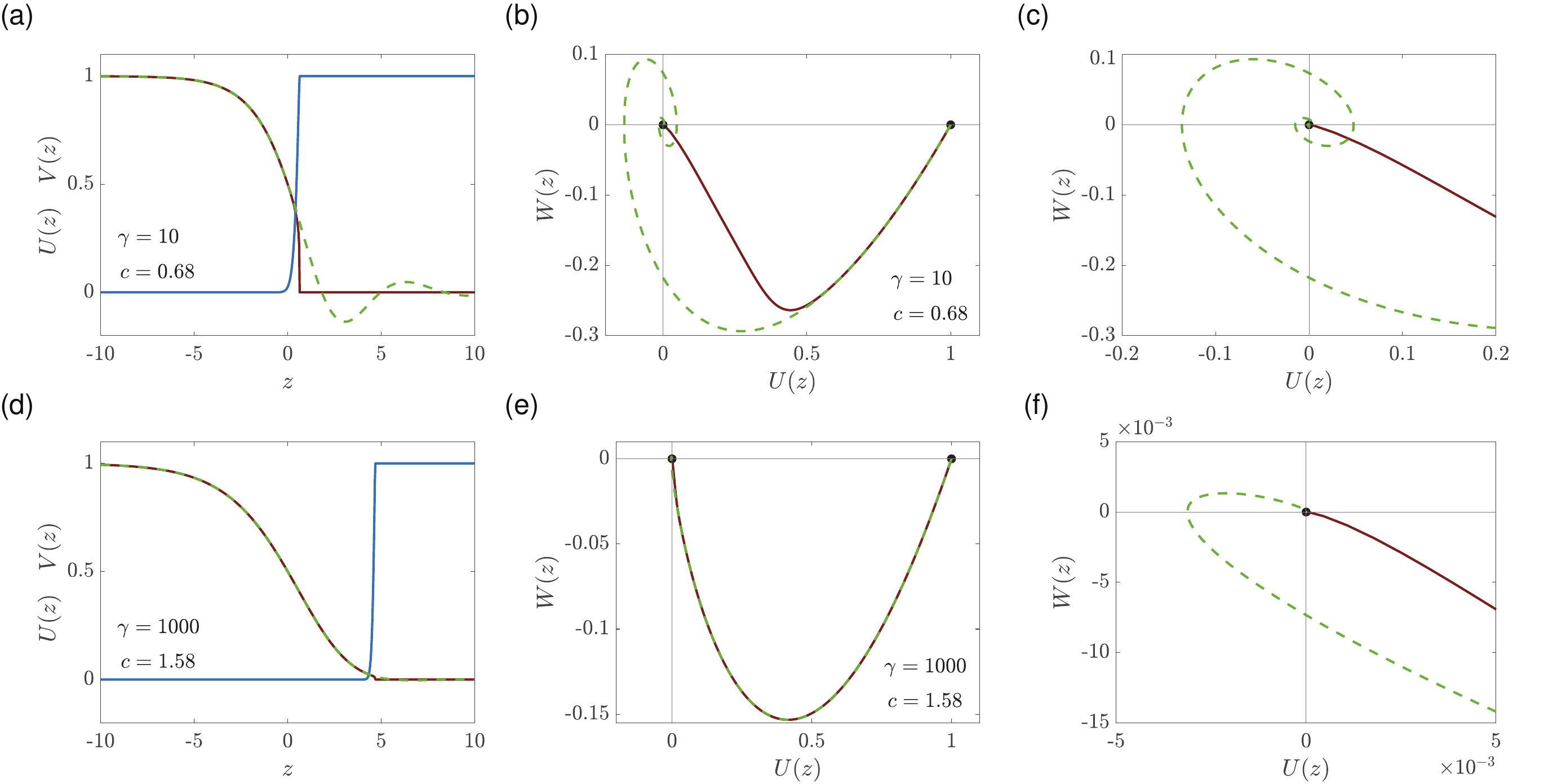}
		\caption{\textbf{Travelling wave solutions of the invasion model with $\gamma \gg 1$ and $\mathcal{V} = 1$.}  (a) and (d) show late-time numerical solutions of Equations (\ref{eq:GouvdiffUNonDimensional})--(\ref{eq:GouvdiffVNonDimensional}) with $\gamma=10$ and $\gamma=1000$, respectively.  The solution for $U(z)$ is shown in solid brown and the solution for $V(z)$ is shown in solid blue.  The approximate solution for $U(z)$ obtained from the Fisher-KPP phase plane is shown in dashed green.  (b) and (c) show a projection of the $(U,V,W)$ phase space for the invasion model projected onto the $(U,W)$ plane together with the projection of the three-dimensional heteroclinic orbit in solid brown.  The trajectory from Fisher-KPP phase plane is shown in dashed green.  (c) and (f) show magnified regions near the origin in (b) and (e), respectively.}
		\label{fig:5}
	\end{figure}	
\end{landscape}

\cbl In this section we demonstrate a relationship between the three-dimensional dynamical system and phase space for the invasion model with the far simpler  two-dimensional dynamical system and phase plane for the simpler Fisher-KPP model. \cb.  To motivate this we compare various solutions for  $\gamma = 10$ and $1000$ with $\mathcal{V}=0.5$ in Figure \ref{fig:4}, and a separate set of comparisons are made for $\gamma = 10$ and $1000$ with $\mathcal{V}=1$ in Figure \ref{fig:5}.

Results in Figure \ref{fig:4}(a) and (d) show the long-time numerical solutions of the invasion model  (\ref{eq:GouvdiffUNonDimensional})--(\ref{eq:GouvdiffVNonDimensional}) with $\gamma=10$ and $\gamma= 1000$, respectively.  Comparing the shapes of these two travelling wave solutions confirms that the width of region \RomanNumeralCaps 2 decreases with  $\gamma$.  In particular, the travelling wave profile in Figure \ref{fig:4}(d) confirms that the overlap between the invading cancer density and the retreating skin density is barely noticeable at this scale.  These travelling wave profiles in Figure \ref{fig:4}(a) and (d) are first generated and plotted in the three-dimensional $(U,V,W)$ phase space, and a projection of this phase space and the trajectory is plotted in the $(U,W)$ plane.  This projection looks like a two-dimensional heteroclinic orbit between $(\bar{U},\bar{W}) = (1,0)$ and $(\bar{U},\bar{W}) = (0,0)$.  To make the connection with the simpler Fisher-KPP model explicit, we superimpose a numerical trajectory obtained from the Fisher-KPP phase plane (\ref{eq:ODEdUdzFKPP})--(\ref{eq:ODEdWdzFKPP}) for the appropriate value of $c$ obtained from the long-time numerical solutions of (\ref{eq:GouvdiffUNonDimensional})--(\ref{eq:GouvdiffVNonDimensional}).  In both cases we see that the projection of the trajectory associated with the invasion model and the trajectory associated with the simpler Fisher-KPP  model compare very well, particularly in Figure \ref{fig:4}(e) where $\gamma=1000$.  The main discrepancy between the trajectories is near the origin.  Additional comparisons in Figure \ref{fig:4}(c) and (f) to show details of the trajectories near the origin where the differences are clear.  Indeed, in both cases we see that the trajectory for the simpler Fisher-KPP  model spirals into the origin, whereas the trajectory for the invasion model does not~\cite{Murray02}.

Overall, the results in Figure \ref{fig:4} reveal a novel application of the well-known phase plane for travelling wave solutions of the Fisher-KPP model since we show that trajectories in this phase plane provide a good approximation to projections of the three-dimensional trajectories associated with travelling wave solutions of the more complicated invasion model (\ref{eq:GouvdiffUNonDimensional})--(\ref{eq:GouvdiffVNonDimensional}). \cbl  This comparison is mathematically interesting since standard phase plane analysis of travelling wave solutions to the Fisher-KPP model are limited to $c \ge 2$, given, the heteroclinic  trajectories for $c < 2$ are disregarded on physical grounds~\cite{Murray02}, but here we find that these trajectories provide a good approximation for the shape of travelling wave solutions to the more complicated invasion model. \cb To highlight the value of this approximation, we take the $(U,W)$ trajectory from the Fisher-KPP phase plane and plot the profile as a function of $z$ in Figure \ref{fig:4}(a) and (d), where we see that the profile obtained from the simpler Fisher-KPP model approximates the shape of travelling wave profile for the full invasion model.  This approximation is particularly accurate in Figure \ref{fig:4}(d) for $\gamma=1000$.  In contrast, while the Fisher-KPP approximation in Figure \ref{fig:4}(a) is quite reasonable where $z < 2$, it is relatively poor in the region $2 < z < 5$ because of the more pronounced oscillation.

Results in Figure \ref{fig:5} are presented in the exact same format as the results in Figure \ref{fig:4}, except that here we have $\mathcal{V}=1$ so we have sharp-fronted travelling wave solutions.  Comparing the shape of the travelling wave solutions in Figure \ref{fig:5}(a) and (d) again confirms that the width of the overlap region decreases with $\gamma$.  The projections of the three-dimensional phase space onto the $(U,W)$ plane in Figure \ref{fig:5}(b) and (e) take the form of a two-dimensional heteroclinic orbit between  $(\bar{U},\bar{W}) = (1,0)$ and $(\bar{U},\bar{W}) = (0,0)$.  Comparing the projections of the three-dimensional trajectory with the numerical trajectory obtained from the Fisher-KPP model (\ref{eq:ODEdUdzFKPP})--(\ref{eq:ODEdWdzFKPP}), confirms that the simpler dynamical system provides a good approximation to the three-dimensional dynamical system when $\gamma$ is large, with only small discrepancies near the origin, as shown in Figure \ref{fig:5}(c) and (f).  In this case, comparing the $U(z)$ profiles in Figure \ref{fig:5}(d) shows that the entire shape of the travelling wave is approximated very well when $\gamma=1000$, but we see a more clear discrepancy in Figure \ref{fig:5}(a) for $\gamma=10$ since this leads to $c=0.68$ and more pronounced oscillations about $U=0$ at the front of the travelling wave.

\subsection{Slow decay: $\gamma \ll 1$}
We now turn our attention to the limit where cancer cells consume skin cells very slowly, $\gamma \ll 1$.  In this limit we find that it is necessary to treat the cases $\mathcal{V} < 1$ and $\mathcal{V} = 1$ separately, as we will now illustrate.

\subsubsection{$\gamma \ll 1$ and $\mathcal{V} < 1$}\label{sec:GammaInfinityVsmallerThan1}

Preliminary numerical results in Figure \ref{fig:2}(a)--(d) show that the width of region \RomanNumeralCaps 2, the overlap region,  increases as $\gamma \rightarrow 0$.  To analyse this behaviour we seek a perturbation solution by treating $\gamma$ as a small parameter, and it is useful to recall from Section \ref{sec:ModelPreliminaryResults} that when $\mathcal{V} < 1$ we have $c = c_{\textrm{min}} = 2(1-\mathcal{V})$ for $\gamma < \gamma_{\textrm{c}}$ so that we take $c= c_{\textrm{min}}$ in our small $\gamma$ analysis.  Re-scaling the dependent variable $\tilde{z} = \gamma z$ gives
\begin{align}
\label{eq:ODEUztilde}
\gamma^2\dfrac{\mathrm{d}}{\mathrm{d} \tilde{z}}\left[(1-V)\dfrac{\mathrm{d}U}{\mathrm{d}\tilde{z}}\right] +c_{\textrm{min}} \gamma \dfrac{\mathrm{d} U}{\mathrm{d} \tilde{z}} + U(1-U-V) &= 0, &-\infty < \tilde{z} < \infty, \\
c_{\textrm{min}}\dfrac{\mathrm{d} V}{\mathrm{d} \tilde{z}}- U V &= 0, &-\infty < \tilde{z} < \infty. \label{eq:ODEVztilde}
\end{align}

We now seek perturbation solutions of the form
\begin{align}
U(\tilde{z}) &=U_0(\tilde{z})+\gamma U_1(\tilde{z})+\mathcal{O}(\gamma^2),\label{eq:ExpansionUztilde}\\
V(\tilde{z}) &=V_0(\tilde{z})+\gamma V_1(\tilde{z})+\mathcal{O}(\gamma^2).\label{eq:ExpansionVztilde}
\end{align}
Substituting these expansions into Equation (\ref{eq:ODEUztilde}) shows that we have $U_0(1-U_0-V_0)=0$ at leading order, so that $V_0(\tilde{z})+U_0(\tilde{z}) = 1$. The differential equations governing the terms in the perturbation solution are therefore given by
\begin{align}
c_{\textrm{min}} \dfrac{\mathrm{d} U_0}{\mathrm{d} \tilde{z}} + U_0 (1-U_0) &= 0, \label{eq:BVP_U0ztilde}\\
c_{\textrm{min}} \dfrac{\mathrm{d} U_0}{\mathrm{d} \tilde{z}} - U_0(U_1+V_1) &= 0, \label{eq:BVP_U1ztilde}\\
c_{\textrm{min}} \dfrac{\mathrm{d} V_1}{\mathrm{d} \tilde{z}} -  V_1( 2U_0-1) + V_0^2 &=0.\label{eq:BVP_V1ztilde}
\end{align}
with boundary conditions $U_0 \rightarrow 1,  V_0 \rightarrow 0, U_1 \rightarrow 0, V_1 \rightarrow 0$ as $\tilde{z} \rightarrow  -\infty$, $U_0 \rightarrow 0, V_0 \rightarrow \mathcal{V}, U_1 \rightarrow 0, V_1 \rightarrow 0$ as $\tilde{z} \rightarrow \infty$.

We solve these differential equations using the following strategy.   Equation (\ref{eq:BVP_U0ztilde}) can be solved for $U_0(\tilde{z})$ directly using separation of variables, and from this we can evaluate $V_0(\tilde{z}) = 1 - U_0(\tilde{z})$. Given the $\mathcal{O}(1)$ solutions, we simply rearrange Equation (\ref{eq:BVP_U1ztilde}) to obtain $U_1(z)$ directly, and the solution of Equation (\ref{eq:BVP_U1ztilde}) is obtained using an integrating factor.  In summary, the solutions of these differential equations are
\begin{align}
U_0(\tilde{z}) &= \dfrac{1}{1+\left(\dfrac{\mathcal{V}}{1-\mathcal{V}}\right)\exp{\left(\dfrac{\tilde{z}}{c_{\textrm{min}}}\right)}}, \quad
V_0(\tilde{z}) = \dfrac{1}{1+\left(\dfrac{1-\mathcal{V}}{\mathcal{V}}\right)\exp{\left(-\dfrac{\tilde{z}}{c_{\textrm{min}}}\right)}},\label{eq:U1ztilde}\\
U_1(\tilde{z}) &= \dfrac{-\mathcal{V} \exp{\dfrac{\tilde{z}}{c_{\textrm{min}}}}}{\left(\mathcal{V} \left[\exp{\left(\dfrac{\tilde{z}}{c_{\textrm{min}}}\right)}-1\right]+1\right)^2}, \quad
V_1(\tilde{z}) = \dfrac{\mathcal{V}^2\left[1-\exp{\left(\dfrac{\tilde{z}}{c_{\textrm{min}}}\right)}\right] \exp{\left(-\dfrac{\tilde{z}}{c_{\textrm{min}}}\right)}}{\left[\exp{\left(-\dfrac{\tilde{z}}{c_{\textrm{min}}}\right)}(\mathcal{V}-1)-\mathcal{V}\right]^2}. \label{eq:V1ztilde}
\end{align}
where we have evaluated the constants of integration by setting $V(0)=\mathcal{V}/2$.

\begin{landscape}
	\begin{figure}[h!]
		\centering
		\includegraphics[height=1\textheight]{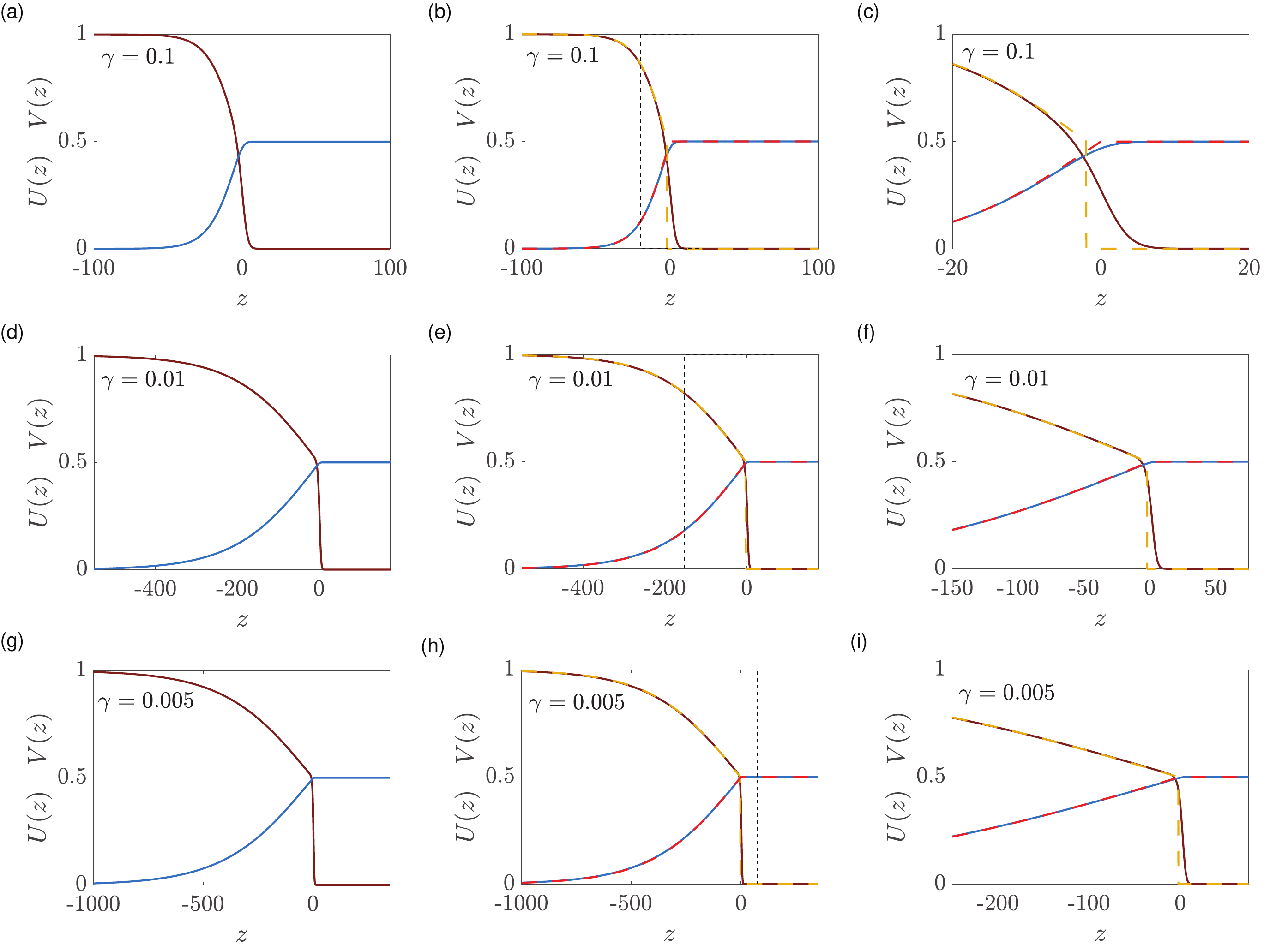}
		\caption{\textbf{Comparison of PDE solutions with perturbation solution when $\gamma \ll 1$ and $\mathcal{V} = 0.5$.} Travelling wave solutions $U(z)$ and  $V(z)$ obtained from PDE with $\Delta x = \Delta t = 1 \times 10^{-2}$, $\gamma = 0.1$ in (a)--(c), $\gamma = 0.01$ in (d)--(f) and $\gamma = 0.005$ in (g)--(i) are illustrated in solid brown and solid blue for $U(z)$ and $V(z)$ respectively. $\mathcal{O}(\varepsilon)$ perturbation solution is illustrated in dashed yellow respectively for $U(z)$ and in dashed red for $V(z)$. Solutions shown in (c),(f) and (i) are magnification of region of interest in (b),(e) and (h) respectively.}
		\label{fig:6}
	\end{figure}
\end{landscape}

Results in Figure \ref{fig:6} show long-time solutions of  (\ref{eq:GouvdiffUNonDimensional})--(\ref{eq:GouvdiffVNonDimensional}) in (a), (d) and (g) for $\gamma=0.1, 0.01$ and $0.005$, respectively.  Comparing the shapes of these travelling waves confirms that the width of the overlap region increases as $\gamma$ decreases.  Results in Figure \ref{fig:6}(b), (e) and (h) compare the shape of the late-time PDE solutions with the $\mathcal{O}(\gamma)$ perturbation solutions, and we see that the accuracy of the approximate perturbation solutions improves as $\gamma$ decreases, as expected.  Results in Figure \ref{fig:6}(c), (f) and (i) compare the numerical solutions and the perturbation solutions within the regions highlighted by the dashed rectangles in Figure \ref{fig:6}(b), (e) and (h). Again, we see the accuracy of the perturbation solution increases as $\gamma$ decreases, and the perturbation solution captures the sharp transition region reasonably accurately as $\gamma \to 0$.

\subsection{$\gamma \ll 1$ and $\mathcal{V} = 1$} \label{sec:GammaInfinityVequal1}
To analyse the shape of the travelling wave for $\gamma \ll 1$ with $\mathcal{V}=1$ we consider the desingularised system
\begin{align}
\dfrac{\mathrm{d}^2 U}{\mathrm{d} \zeta^2} + c\dfrac{\mathrm{d} U}{\mathrm{d} \zeta} + U(1-V)(1-U-V)&=0, \label{eq:ODEUzeta}\\
\dfrac{\mathrm{d} V}{\mathrm{d} \zeta} - \dfrac{\gamma U V}{c} (1-V) &=0 , \label{eq:ODEVzeta}
\end{align}
with boundary conditions $U \rightarrow 1,  V \rightarrow 0$ as $\zeta \rightarrow  -\infty$, $U \rightarrow 0, V \rightarrow 1$ as $\zeta \rightarrow  \infty$.

Numerical results in Figure \ref{fig:2} show that when $\mathcal{V} = 1$ we have $c\rightarrow 0$ as $\gamma\rightarrow 0$, so we write  $c=\tilde{c}\gamma$ so that $\tilde{c}$ is $\mathcal{O}(1)$.  Like in the previous section, we re-scale the independent variable $\tilde{\zeta} = \gamma \zeta$ to give
\begin{align}
\label{eq:ODEdUdzetatilde}
\dfrac{\mathrm{d}^2 U}{\mathrm{d} \tilde{\zeta}^2}+\tilde{c}\gamma\dfrac{\mathrm{d} U}{\mathrm{d} \tilde{\zeta}} + U(1-U-V)(1-V) &= 0, &-\infty<\tilde{\zeta}<\infty\\
\tilde{c}\dfrac{\mathrm{d} V}{\mathrm{d} \tilde{\zeta}}- U V (1-V) &= 0, &-\infty<\tilde{\zeta}<\infty,\label{eq:ODEdVdzetatilde}
\end{align}
with boundary conditions $U \rightarrow 1,  V \rightarrow 0$ as $\tilde{\zeta} \rightarrow  -\infty$, $U \rightarrow 0, V \rightarrow 1$ as $\tilde{\zeta} \rightarrow  \infty$.

Seeking a perturbation solution of the form
\begin{align}
U(\tilde{\zeta}) &=U_0(\tilde{\zeta})+\gamma U_1(\tilde{\zeta})+\mathcal{O}(\gamma^2),\label{eq:ExpansionUzetatilde}\\
V(\tilde{\zeta}) &=V_0(\tilde{\zeta})+\gamma V_1(\tilde{\zeta})+\mathcal{O}(\gamma^2),\label{eq:ExpansionVzetatilde}
\end{align}
leads to a system of coupled differential equations for the leading order terms,
\begin{align}
\label{eq:BVP_U0zetatilde}
\dfrac{\mathrm{d}^2 U_0}{\mathrm{d} \tilde{\zeta}^2}+ U_0(1-U_0-V_0)(1-V_0) &= 0,\\
\tilde{c}\dfrac{\mathrm{d} V_0}{\mathrm{d} \tilde{\zeta}}- U_0 V_0 (1-V_0) &= 0,\label{eq:BVP_V0zetatilde}
\end{align}
with boundary conditions $U_0 \rightarrow 1,  V_0 \rightarrow 0$ as $\tilde{\zeta} \rightarrow -\infty$, $U_0 \rightarrow 0, V_0 \rightarrow 1$ as $\tilde{\zeta} \rightarrow \infty$.  Unfortunately we are unable to obtain a closed-form solution of Equations (\ref{eq:BVP_U0zetatilde})--(\ref{eq:BVP_V0zetatilde}) and we do not proceed further with this approach.

\section{Fast travelling waves, $c > c_{\textrm{min}}$}\label{sec:CGreaterThanCmin}
\begin{figure}[H]
	\centering
	\includegraphics[width=1\linewidth]{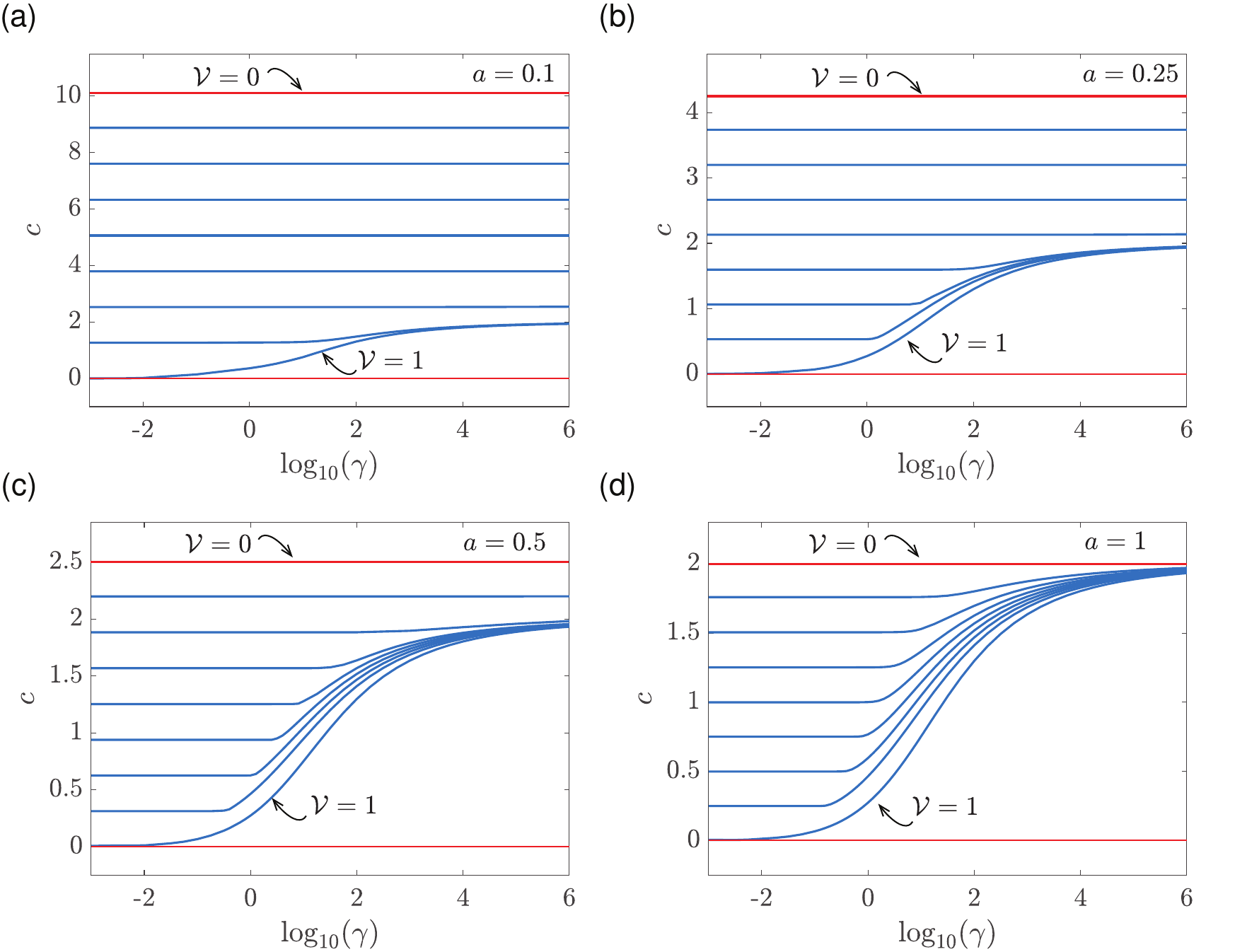}
	\caption{\textbf{Relationship between $c$, $\gamma$ and $\mathcal{V}$ for slowly decaying initial $u(x,0)$.} Numerical estimates of $c$ are obtained from long-time solutions of Equations (\ref{eq:GouvdiffUNonDimensional})--(\ref{eq:GouvdiffVNonDimensional}) with the initial condition given by Equations (\ref{eq:ICUNonCompactSupport})--(\ref{eq:ICVNonCompactSupport}).  Results are presented with  $\beta = 10$ and $a = 0.1, 0.25, 0.5$ and $1$, as indicated.  Time-dependent PDE solutions are obtained using $\Delta x = \Delta t= 1 \times 10^{-2}$, for $1\times10^{-3}\le \gamma \le 1\times10^{6}$ and $\mathcal{V} = 1/8, 2/8, 3/8, 4/8, 5/8, 6/8, 7/8$ and $1$.}
	\label{fig:7}
\end{figure}

Our focus so far has been on the long-time limit of the time-dependent solutions of (\ref{eq:GouvdiffUNonDimensional})--(\ref{eq:GouvdiffVNonDimensional}) with initial conditions given by
Equations (\ref{eq:ICUCompactSupport})--(\ref{eq:ICVCompactSupport}) so that $u(x,0)$ has compact support.  This leads to travelling wave solutions with the minimum wave speed, $c_{\textrm{min}}$.  We now examine travelling wave solutions of the same model but with initial conditions given by Equations (\ref{eq:ICUNonCompactSupport})--(\ref{eq:ICVNonCompactSupport}), where $a>0$ controls the far-field decay rate of $u(x,0)$. Results in Figure \ref{fig:7} summarise the numerically-observed travelling wave speed, $c$, for $a=0.1, 0.25, 0.5$ and $1$, as a function of $\mathcal{V}$ and $\gamma$ as indicated. Comparing these results with those in Figure \ref{fig:3} for initial conditions with compact support, we see that some general features of the relationship between $c$, $\mathcal{V}$ and $\gamma$ are maintained, while other features are different.  In general we see that $c$ is a decreasing function of $a$, and that all results suggest that $c$ is  independent of $\gamma$ for sufficiently small $\gamma$ when $\mathcal{V} < 1$.  As $\gamma$ increases, we see that $c$ increases with $\gamma$, but that the limiting value of $c$ as $\gamma \to \infty$ depends upon the decay rate, $a$.  Careful comparison of the results in Figure \ref{fig:7}(d) for $a=1$ shows that different choices of $\gamma$ and $\mathcal{V}$ lead to the exact same travelling wave speed as in Figure \ref{fig:3} for initial conditions with compact support, which can be thought of as letting $a \to \infty$ in (\ref{eq:ICUNonCompactSupport})--(\ref{eq:ICVNonCompactSupport}).  We will now explain some of these observed trends analytically.

To understand the relationship between the decay rate of the initial condition and the asymptotic wave speed, $c$, we examine the leading edge of the travelling wave where $u \ll 1$, giving
\begin{align}
&\dfrac{\partial \tilde{u}}{\partial t}= (1-\tilde{v})\dfrac{\partial^2 \tilde{u}}{\partial x^2} + (1-\tilde{v}) \tilde{u}, \label{eq:LinearULargex}\\
&\dfrac{\partial \tilde{v}}{\partial t}=-\gamma \tilde{u}\tilde{v}.\label{eq:LinearVLargex}
\end{align}
Assuming the travelling wave solution takes the form $\tilde{u} \sim \textrm{exp}[-a (x-ct)]$ for large $x$, substituting this into  Equation (\ref{eq:LinearULargex}) gives
\begin{equation}
c =  (1-\mathcal{V}) \left(a + \dfrac{1}{a}\right). \label{eq:DispersionRelationship}
\end{equation}
This dispersion relationship is similar to the analogous result for the Fisher-KPP model~\cite{Murray02}.  This simple relationship explains some of the observations in Figure \ref{fig:7} where we see that $c$ is independent of $\gamma$ for $\gamma < \gamma_\textrm{c}$.  Indeed, this constant speed is given by Equation (\ref{eq:DispersionRelationship}).  Unfortunately, this dispersion relationship does not provide any insight into the relationship between $c$, $\mathcal{V}$ and $\gamma$ for $\gamma > \gamma_{\textrm{c}}$, nor any insight into the shape of the resulting travelling waves.

To address the shape of the travelling waves for $c > c_{\textrm{min}}$ we extend the approach of Canosa \cite{Canosa1973} who noted that travelling wave solutions of the Fisher-KPP model become increasingly wide as $c \to \infty$.  Our numerical simulation results suggest that travelling wave solutions of the invasion model behave similarly, so we explore this behaviour by re-scaling the independent variable, $\overline{z} = z/c$, giving
\begin{align}
\label{eq:ODEUzetaBigC}
\dfrac{\mathrm{d} U}{\mathrm{d} \overline{z}} + \frac{1}{c^2}\dfrac{\mathrm{d}}{\mathrm{d} \overline{z} }\left[(1-V)\dfrac{\mathrm{d}U}{\mathrm{d} \overline{z}}\right]  + U(1-U-V) &= 0,  \quad  -\infty < \overline{z} < \infty,\\
\dfrac{\mathrm{d} V}{\mathrm{d} \overline{z}}-\gamma U V &= 0,  \quad  -\infty < \overline{z} < \infty, \label{eq:ODEVzetaBigC}
\end{align}
with boundary conditions $U(\overline{z}) \rightarrow 1$ and $V(\overline{z}) \rightarrow 0$ as $\overline{z} \rightarrow -\infty$,  and $U(\overline{z}) \rightarrow 0$ and $V(\overline{z}) \rightarrow \mathcal{V}$ as  $\overline{z} \rightarrow \infty$. In this re-scaled coordinate we seek perturbation solutions in terms of the small parameter $\varepsilon = 1/c^2$,
\begin{align}
U(\overline{z}) &= U_0(\overline{z}) + \varepsilon U_1(\overline{z}) + \mathcal{O}(\varepsilon^2),\label{eq:ExpansionUBigC}\\
V(\overline{z}) &= V_0(\overline{z}) + \varepsilon V_1(\overline{z}) + \mathcal{O}(\varepsilon^2).\label{eq:ExpansionVBigC}
\end{align}
Substituting Equations (\ref{eq:ExpansionUBigC})--(\ref{eq:ExpansionVBigC}) into (\ref{eq:ODEUzetaBigC})--(\ref{eq:ODEVzetaBigC}) leads to
\begin{align}
\dfrac{\mathrm{d} U_0}{\mathrm{d} \overline{z}}  + U_0(1-U_0-V_0) &= 0, \label{eq:BVP_U0BigC} \\
\dfrac{\mathrm{d} V_0}{\mathrm{d} \overline{z}}-\gamma U_0 V_0 &= 0,  \label{eq:BVP_V0BigC}
\end{align}
with boundary conditions $U_0(\overline{z}) \rightarrow 1$ and $V_0(\overline{z}) \rightarrow 0$ as $\overline{z} \rightarrow -\infty$, $U_0(\overline{z}) \rightarrow 0$ and $V_0(\overline{z}) \rightarrow \mathcal{V}$ as $\overline{z} \rightarrow \infty$. Unlike Canosa~\cite{Canosa1973}, these differential equations for the $\mathcal{O}(1)$ terms in the perturbation solution do not have a closed-form solution.  Nevertheless, we make progress by re-writing Equations (\ref{eq:BVP_U0BigC})--(\ref{eq:BVP_V0BigC}) as
\begin{equation}
\dfrac{\mathrm{d} U_0}{\mathrm{d} V_0} = -\dfrac{1-U_0-V_0}{\gamma V_0}, \label{eq:ODEdiffUVBigC}
\end{equation}
with $U_0(\mathcal{V})= 0$.  The solution of this problem is given by
\begin{equation}
U_0(V_0) = \dfrac{1}{(\gamma-1)} \left[\left(V_0+\gamma-1\right) -({\mathcal{V}}+\gamma-1) \left(\dfrac{V_0}{\mathcal{V}}\right)^{\left(\dfrac{1}{\gamma}\right)}\right]. \label{eq:solutionU0V0BigC}
\end{equation}
For the special case $\gamma=1$, this solution can be written as
\begin{equation}
U_0(V_0) = 1  +\left( \dfrac{V_0}{\mathcal{V}} \right) \left(\mathcal{V} \log \left[\dfrac{V_0}{\mathcal{V}}\right] -1 \right).
\end{equation}

\begin{figure}[H]
	\centering
	\includegraphics[width=0.9\linewidth]{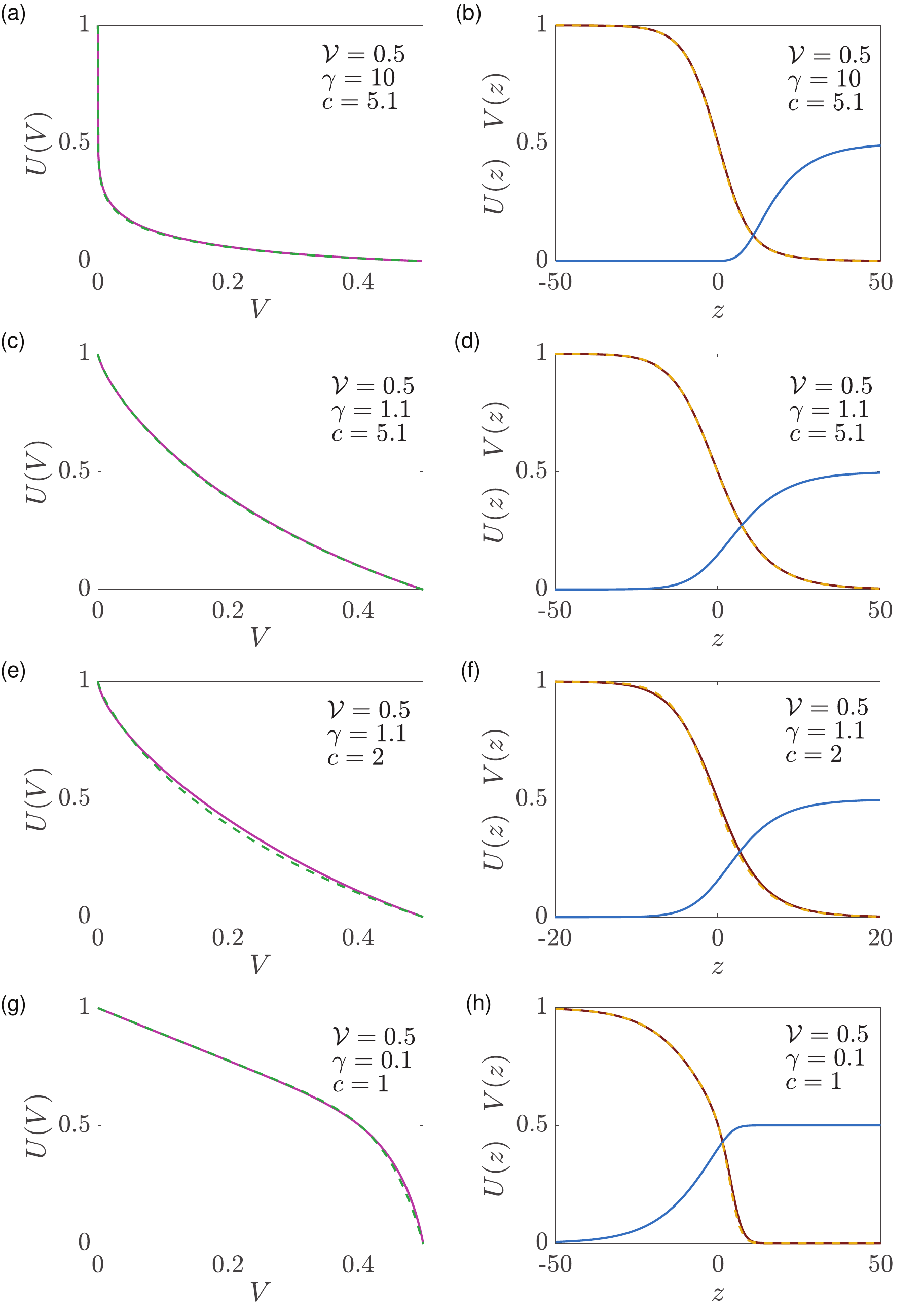}
	\caption{\textbf{Comparison of numerical and perturbation solutions for $c \ge c_\textrm{cmin}$.} The $\mathcal{O}(1)$ perturbation solution shown in dashed green is compared to solution $U(V)$ from PDE shown in solid pink, in (a), (c), (e) and (g). Solutions $U(z)$ and $V(z)$ from PDE are shown in solid brown and in solid blue respectively in (b), (d), (f) and (h). Perturbation solution is shown in dashed yellow as $U(z)$ by using $V_0=V(z)$ from PDE solution. The initial conditions (\ref{eq:ICUNonCompactSupport})--(\ref{eq:ICVNonCompactSupport}) with $\beta = 10$ are used in (a)--(d), where $a = 0.1$, and in (e)--(f) where $a = 0.27$. The initial conditions with compact support are used in (g)--(h).}
\label{fig:8}
\end{figure}

Results in Figure \ref{fig:8} compare travelling wave solutions with various $c$ with the $\mathcal{O}(1)$ perturbation solution (\ref{eq:solutionU0V0BigC}).  Results in the left column of Figure \ref{fig:8} show $U$ as a function of $V$, as explicitly defined by Equation (\ref{eq:solutionU0V0BigC}) superimposed on curves obtained from long-time numerical solutions of Equations (\ref{eq:GouvdiffUNonDimensional})--(\ref{eq:GouvdiffVNonDimensional}) that are plotted in the same format.  Results in the right column of Figure \ref{fig:8} show the late time solutions of Equations (\ref{eq:GouvdiffUNonDimensional})--(\ref{eq:GouvdiffVNonDimensional}) plotted in the travelling wave coordinate, $z$, superimposed with the perturbation solution.   Results in Figure \ref{fig:8}(a)--(d) show perturbation results for $c = 5.1$ and different choices of $\gamma$.  At this scale the perturbation solution is visually indistinguishable from the numerical solutions. Results in Figure \ref{fig:8}(e)--(f) show that the perturbation solution performs well for $c=2$, but that we can begin to see some discrepancy between the perturbation and numerical results.  Interestingly, the perturbation solution in Figure \ref{fig:8}(g)--(h) leads to reasonably accurate solutions for $c=1$, despite the fact that Equation (\ref{eq:solutionU0V0BigC}) is valid in the limit $c \to \infty$.

\section{Conclusion and Outlook}
In this work we study travelling wave solutions of a model of cellular invasion, (\ref{eq:GouvdiffUNonDimensional})--(\ref{eq:GouvdiffVNonDimensional}), where the migration and proliferation of the invasive population is coupled to the degradation of surrounding skin tissues~\cite{Browning2019}.  Time-dependent numerical solutions of the governing PDEs show that there is a complicated relationship between the travelling wave speed $c$ and: (i) $\gamma$,  the rate at which the invasive population degrades the surrounding skin tissues; and, (ii) $\mathcal{V}$, the far field density of surrounding tissues.  Numerical exploration shows that long-time travelling wave solutions are smooth without compact support for $\mathcal{V} < 1$, or sharp-fronted with compact support with $\mathcal{V}=1$.  The relationship between $c$ and the parameters in the model are partially established in this work.  Numerical simulations and phase space analysis show that for initial conditions with compact support, we have  $c = 2 (1- \mathcal{V})$, which is independent of $\gamma$ for $\gamma < \gamma_\textrm{c}$. Further numerical simulations show that $c$ increases with $\gamma$ for $\gamma > \gamma_\textrm{c}$, with $c \to 2^-$ as $\gamma \to \infty$, but the precise details of this relationship are not revealed through standard phase space analysis.  Of great interest is the fact that we always have $c < 2$ for the invasion model for initial conditions with compact support; this is very different to the standard Fisher-KPP model where $c \ge 2$.  We also show that time-dependent PDE solutions for initial conditions without compact support lead to travelling wave solutions with larger wave speed.

Analysis of the invasion model for fast decay, $\gamma \gg 1$, indicates that the width of the overlap region decreases with $\gamma$.  This means that the density of the invading population becomes uncoupled from the density of the surrounding skin tissues, and suggests that the shape of the invading density profile is related to the shape of the travelling wave solution of the Fisher-KPP model.  This is intriguing since the invasion model is associated with travelling waves with $c < 2$ whereas the Fisher-KPP model is associated with $c \ge 2$.  Indeed, our results show that the well-known phase plane associated with travelling wave solutions of the Fisher-KPP model provides a novel approximation to the shape of the travelling wave solution of the invasion model for fast decay.  This observation is mathematically interesting because standard analysis of the Fisher-KPP model disregards the phase plane for $c < 2$~\cite{Murray02}, whereas here we find that this previously disregarded phase plane is closely related to our model of invasion.

\cbl Overall, our analysis and numerical exploration shows how a simple mathematical model of invasion can generate biological hypotheses that can be further studied experimentally or clinically.  For example, our model predicts that when the cancer population shares the exact same carrying capacity that the normal tissue and $\mathcal{V}$=1, the resulting invasion front is sharp and has compact support. Conversely,  when the cancer population has a different carrying capacity and can grow to a larger density than the surrounding tissues the resulting invasion is smooth and without compact support.  The differences between invasion fronts having compact support or being smooth is very important when considering surgical intervention, since it is always possible to completely remove an invasive population with compact support by excising tissue ahead of the invading front.  In contrast, it is theoretically impossible to completely remove an invasive population by excising tissue when the front is smooth and without compact support. \cb

Our observation that the normally-disregarded phase plane associated with travelling wave solutions of the Fisher-KPP model for $c < 2$  can be used to approximate the shape of the travelling wave solutions of the invasion model leads us to an interesting and previously unnoticed link with a very different type of mathematical model of invasion, the Fisher-Stefan model~\cite{Du2010,Du2011,Du2012,Du2014,ElHachem2019,ElHachem2020,ElHachem2021,ElHachem2021b}.  The Fisher-Stefan model involves studying the Fisher-KPP model on a moving boundary, $0 < x < L(t)$.  In this model the density vanishes on the moving boundary, $u(x,L(t))=0$, and the speed of the moving boundary is driven by a one-phase Stefan condition, $\textrm{d} L(t) / \textrm{d}t = -\kappa \partial u(L(t),t)/\partial x$.  Here, $\kappa > 0$ is a constant that relates the speed of the moving boundary to the spatial gradient of density at the moving boundary.   For both the invasion model and the Fisher-Stefan model it has been shown that time-dependent PDE solutions eventually evolve to travelling wave solutions with $c < 2$~\cite{Du2010,Du2011,Du2012,Du2014,ElHachem2019,ElHachem2020,ElHachem2021,ElHachem2021b} and the shape of the invading profiles in both cases is given by the normally-disregarded phase plane of the well-known Fisher-KPP model.  This is very interesting because the normally-disregarded phase plane trajectories imply $U(z) < 0$ for certain intervals in $z$.  However, here and in the Fisher-Stefan model, the profile of interest is given by a truncated trajectory in the phase plane where $U(z)>0$.

There are many opportunities to extend the work presented in this study.  There are several assumptions in the mathematical model (\ref{eq:GouvdiffUNonDimensional})--(\ref{eq:GouvdiffVNonDimensional}), that could be relaxed or varied.  Such extensions could involve working with different nonlinear diffusivity functions in (\ref{eq:GouvdiffUNonDimensional}), such as a power law~\cite{McCue2019}.  Another extension of interest would be to explore the impact of using a different nonlinear source term in (\ref{eq:GouvdiffUNonDimensional}) to model the proliferation of cells~\cite{Sanchez1994}.  One of the limitations of our study is that we have not been able to arrive at a mathematical expression for $\gamma_\textrm{c}$, and it would be of great interest to arrive at some approximate expression for this critical decay rate, or to place some bound on that value.  Further extensions could involve working in a different geometry since models of melanoma invasion in both two and three-dimensions with radial symmetry are of great interest for studying malignant invasion, e.g~\cite{Ward1997,Ward1999,Jin2021}. \cbl Finally, we acknowledge that all analysis here is limited to dealing with a continuum mathematical model only.  One of the limitations of working within a continuum framework is that it ignores the role of  stochasticity. An alternative approach to study invasion is to consider individual based stochastic models, e.g. ~\cite{Deutsch2005,Haridas2018b}, which explicitly describe variations between individual cells in the population. \cb

\appendix
\section{Numerical methods}

We solve Equations (\ref{eq:GouvdiffUNonDimensional})--(\ref{eq:GouvdiffVNonDimensional}) on $0 < x < L$ by uniformly discretising the domain with $N$ mesh points, with spacing $\Delta x$.  We approximate the spatial derivatives in Equations (\ref{eq:GouvdiffUNonDimensional})--(\ref{eq:GouvdiffVNonDimensional}) using a central difference approximation, and solve the resulting system of coupled ordinary differential equation through time using an implicit Euler approximation, giving
\begin{align}
\dfrac{u_i^{j+1} - u_i^{j}}{\Delta t} = & \dfrac{1}{2\Delta x^2}\bigg(\left[2 - \left(v_{i+1}^{j+1} + v_i^{j+1}\right) \right]\left( u_{i+1}^{j+1} - u_i^{j+1} \right) - \left[2 - \left(v_i^{j+1} + v_{i-1}^{j+1}\right)\right] \left(u_i^{j+1} - u_{i-1}^{j+1}\right)\bigg)  \notag \\
& +   u_i^{j+1} \left(1-u_i^{j+1} - v_i^{j+1}\right), \\
\dfrac{v_i^{j+1} - v_i^{j}}{\Delta t} &= -\gamma u_i^{j+1} v_i^{j+1},
\end{align}
where $\Delta t$ is the time step, $i$ is the spatial finite difference mesh index and $j$ is the temporal index so that $u_{i}^{j} \approx u(x=(i-1)\Delta x, j \Delta t)$. Discretising the boundary conditions for $u(x,t)$ leads to
\begin{equation}
u_{2}^{j+1}-u_{1}^{j+1} = 0,  \quad  u_{N}^{j+1}-u_{N-1}^{j+1} = 0. \label{eq:BoundCondDiscrete}
\end{equation}
Note that there are no boundary conditions for $v$, so the spatial index on the discretisation for $v$ is  $i=1,2,\ldots, N$.  This discretisation leads to a coupled system of nonlinear algebraic equations for $u_i^{j+1}$ and $v_i^{j+1}$, which are solved sequentially to take advantage of the tridiagonal structure of the discretised equations.  The nonlinear equations are solved using Newton-Raphson iterations that continue until the maximum change in the dependent variables falls below some tolerance $\epsilon$ in each time step.  For all problems considered we always check that our choices of $\Delta x$, $\Delta t$ and $\epsilon$ lead to grid-independent results. Matlab software to implement these numerical solutions is available on \href{https://github.com/ProfMJSimpson/Cellular_Invasion_ElHachem_2021}{GitHub}.

To estimate the travelling wave speed from our time-dependent PDE solutions we specify a contour value, $u(x,t) = U$.  At the end of each time step in we use linear interpolation to find the value of $X$ such that $u(X,t) = U$. At the end of each time step we have estimates of both $X(t + \Delta t)$ and $X(t)$, allowing us to estimate the speed at which the contour moves
\begin{equation}\label{eq:NumericalWaveSpeed}
c = \dfrac{X(t + \Delta t)-X(t)}{\Delta t}.
\end{equation}
Evaluating Equation (\ref{eq:NumericalWaveSpeed}) at each time step leads to a time series of estimates of $c$, and we find that these estimates asymptote to some positive constant value as $t$ becomes sufficiently large.  For all results presented we set $U=0.5$, \cbl but we find that our estimates of $c$ are independent of this choice of density contour.\cb

To construct phase planes for Fisher-KPP equation, we solve Equation (\ref{eq:ODEFKPP}) numerically using Heun's method with a constant step size $\textrm{d}z$.  Since we are interested in examining trajectories that leave the saddle $(1,0)$ along the unstable manifold we choose the initial location on the trajectory to be on the appropriate unstable manifold and sufficiently close to $(1,0)$. Matlab software to generate these phase planes and associated trajectories are available on  \href{https://github.com/ProfMJSimpson/Cellular_Invasion_ElHachem_2021}{GitHub}.

\newpage

\noindent
\paragraph{Acknowledgements}  This work is supported by the Australian Research Council (DP200100177). \cbl We thank the two  anonymous referees for helpful suggestions. \cb

\noindent
\paragraph{Contributions} All authors conceived and designed the study; M.El-H. performed numerical and symbolic calculations.  All authors drafted the article and gave final approval for publication.

\noindent
\paragraph{Competing Interests} We have no competing interests.


\begin{thebibliography}{99}
\bibitem{Browning2019} A.P. Browning, P. Haridas, M.J. Simpson. A Bayesian sequential learning framework to parameterise continuum models of melanoma invasion into human skin. Bulletin of Mathematical Biology, 81 (2019) 676--698.

\bibitem{Fisher1937} R.A. Fisher. The wave of advance of advantageous genes. Annals of Eugenics, 7 (1937) 355--369.
	
\bibitem{Kolmogorov1937} A.N. Kolmogorov, P.G. Petrovskii, N.S. Piskunov. A study of the diffusion equation with increase in the amount of substance, and its application to a biological problem. Moscow University Mathematics Bulletin, 1 (1937) 1--26.	

\bibitem{Murray02} J.D. Murray. Mathematical Biology I: An Introduction. Third edition, Springer, New York, (2002).

\bibitem{Sanchez1994} F. S\'{a}nchez Gardu\~{n}o, P.K. Maini. An approximation to a sharp type solution of a density-dependent reaction-diffusion equation. Applied Mathematics Letters, 7 (1994) 47--51.
		
\bibitem{Witelski1994} T.P. Witelski. An asymptotic solution for traveling waves of a nonlinear-diffusion Fisher's equation. Journal of Mathematical Biology, 33 (1994) 1--16.
	
\bibitem{Witelski1995} T.P. Witelski. Merging traveling waves for the porous-Fisher's equation. Applied Mathematics Letters, 8 (1995) 57--62.

\bibitem{McCue2019} S.W. McCue, W. Jin, T.J. Moroney, K-Y. Lo, S-E. Chou, M.J. Simpson. Hole-closing model reveals exponents for nonlinear degenerate diffusivity functions in cell biology. Physica D: Nonlinear Phenomena, 398 (2019) 130--140.

\bibitem{Sherratt90} J.A. Sherratt, J.D. Murray. Models of epidermal wound healing. Proceedings of the Royal Society of London: Series B, 241 (1990) 29--36.

\bibitem{Swanson2003} K.R. Swanson, C. Bridge, J.D. Murray, E.C. Alvord Jr. Virtual and real brain tumors: using mathematical modeling to quantify glioma growth and invasion. Journal of Neurological Sciences, 216 (2003) 1--10.	

\bibitem{Maini2004} P.K. Maini, D.L.S. McElwain, D.I. Leavesley. Traveling wave model to interpret a wound-healing cell migration assay for human peritoneal mesothelial cells. Tissue Engineering, 10 (2004) 475--482.

\bibitem{Sengers2007} B.G. Sengers, C.P. Please, R.O.C. Oreffo. Experimental characterization and computational modelling of two-dimensional cell spreading for skeletal regeneration. Journal of the Royal Society Interface,  4 (2007) 1107--1117.

\bibitem{Gerlee2012} P. Gerlee, S. Nerlander. The impact of phenotypic switching on glioblastoma growth and invasion.  PLoS Computational Biology, 8(6) (2012) e1002556.

\bibitem{Jin2020} W. Jin, K-Y. Lo, Y-S. Sun, YH Ting, M.J. Simpson.  Quantifying the role of different surface coatings in experimental models of wound healing. Chemical Engineering Science 220 (2020) 115609.

\bibitem{Largergren2021} J.H. Lagergren, J.T. Nardini, G. Michael Lavigne, E.M. Rutter, K.B. Flores. Learning partial differential equations for biological transport models from noisy spatio-temporal data. Proceedings of the Royal Society A: Mathematical, Physical and Engineering Sciences, 476 (2021) 20190800.

\bibitem{Largergren2021b} J.H. Largergren, J.T. Nardini, R.E. Baker, M.J. Simpson, K.B. Flores. Biologically-informed neural networks guide mechanistic modelling from sparse experimental data. PLOS Compuational Biology, 16(12): e1008462.

\bibitem{Skellam1951} J.G. Skellam. Random dispersal in theoretical populations. Biometrika, 38 (1951) 196--218.

\bibitem{Lewis1994} M.A. Lewis, P. Kareiva. Allee dynamics and the spread of invading organisms.  Theoretical Population Biology, 43 (1994) 141--158.
	
\bibitem{Holmes1994} E.E. Holmes, M.A. Lewis, J.E. Banks, R.R. Veit. Partial differential equations in ecology: spatial interactions and population dynamics. Ecology, 74 (1994) 17--29.

\bibitem{Shigesada1995} N. Shigesada, K. Kawasaki, Y. Takeda. Modeling stratified diffusion in biological invasions. American Naturalist, 146 (1995) 229--251.

\bibitem{Steel1998} J.  Steele, J. Adams, T. Sluckin. Modelling paleoindian dispersals. World Archaeology, 30 (1998), 286--305.

\bibitem{Kot03} M. Kot. Elements of Mathematical Ecology. Cambridge University Press, Cambridge, (2003).


\bibitem{Levin2003} S.A. Levin, H.C. Muller-Landau, R. Nathan, J. Chave. The ecology and evolution of seed dispersal: a theoretical perspective. Annual Review of Ecology, Evolution, and Systematics. 34 (2003) 575--604.

\bibitem{Canosa1973} J. Canosa. On a nonlinear diffusion equation describing population growth. IBM Journal of Research and Development, 17 (1973) 307--313.
			
\bibitem{Painter2003} K.J. Painter, J.A. Sherratt. Modelling the movement of interacting cell populations.  Journal of Theoretical Biology, 225 (2003) 327--339.	

\bibitem{Byrne2003a} H. Byrne, L. Preziosi. Modelling solid tumour growth using the theory of mixtures.  Mathematical Medicine and Biology: A Journal of the IMA, 20 (2003) 341--366.

\bibitem{Byrne2003b} H.M. Byrne, J.R. King, D.L.S. McElwain, L Preziosi. A two-phase model of solid tumour growth.  Applied Mathematics Letters, 16 (2003) 567--573.
    	
\cbl
\bibitem{Amor2010} D.R. Amor, J. Fort. Virus infection speeds: theory versus experiment. Physical Review E 82, (2010) 061905.

\bibitem{Amor2017} D.R. Amor, R. Monta{\~{n}}ez, S. Duran-Nebreda, R. Sol\'{e}. Spatial dynamics of synthetic microbial mutualists and their parasites. PLoS Computational Biology 13, (2017) e1005689.

\bibitem{Fort2012} J. Fort. Synthesis between demic and cultural diffusion in the Neolithic transition in Europe. Proceedings of the National Academy of Sciences 109, (2012) 18669--18673.

\bibitem{Muller2014} M.J.I. M{\~{u}}ller, I.N. Beverly, D.R. Nelson, A.W. Murray. Genetic drift opposes mutualism during spatial population expansion. Proceedings of the National Academy of Sciences 111, (2014) 1037--1042.
\cb

\bibitem{Haridas2017} P. Haridas, J.A. McGovern, D.L.S. McElwain, M.J. Simpson. Quantitative comparison of the spreading and invasion of radial growth phase and metastatic melanoma cells in a three-dimensional human skin equivalent model. PeerJ, 5 (2017) e3754.
	
\bibitem{Haridas2018} P. Haridas, A.P. Browning, J.A. McGovern, D.L.S. McElwain,  M.J. Simpson. Three-dimensional experiments and individual based simulations show that cell proliferation drives melanoma nest formation in human skin tissue.  BMC Systems Biology, 12 (2018) 34.
		
\bibitem{Gatenby1996} R.A. Gatenby,  E.T. Gawlinski. A reaction-diffusion model of cancer invasion. Cancer Research, 56 (1996) 5745--5753.

\bibitem{Landman1998} K.A. Landman, G.J. Pettet.  Modelling the action of proteinase and inhibitor in tissue invasion. Mathematical Biosciences,  154 (1998) 23--37.

\bibitem{Perumpanani1999}  A.J. Perumpanani, J.A. Sherratt, J. Norbury, H.M. Byrne. A two parameter family of travelling waves with a singular barrier arising from the modelling of extracellular matrix mediated cellular invasion. Physica D: Nonlinear Phenomena, 126 (1999) 145--159.

\bibitem{Marchant2000} B.P. Marchant, J. Norbury, A.J. Perumpanani, Travelling shock waves arising in a model of malignant invasion.  SIAM Journal on Applied Mathematics, 60 (2000) 463--476.

\bibitem{Smallbone2005} K. Smallbone, D.J. Gavaghan, R.A. Gatenby, P.K. Maini. The role of acidity in solid tumour growth and invasion.  Journal of Theoretical Biology, 235 (2005) 476--484.

\bibitem{Landman2008} K.A. Landman, M.J. Simpson, G.J. Pettet,  Tactically-driven nonmonotone travelling waves. Physica D: Nonlinear Phenomena, 237 (2008) 678--691.

\bibitem{Anderson2008} A.R.A. Anderson, V. Quaranta. Integrative mathematical oncology. Nature Reviews Cancer,  8 (2008) 227--234.

\bibitem{Astanin2009} S. Astanin, L. Preziosi. Mathematical modelling of the Warburg effect in tumour cords. Journal of Theoretical Biology, 258 (2009) 578--590.

\bibitem{Fasano2009} A. Fasano, M.A. Herrero, M.R. Rodrigo. Slow and fast invasion waves in a model of acid-mediated tumour growth. Mathematical Biosciences, 220 (2009) 45--56.

\bibitem{Byrne2010} H.M. Byrne. Dissecting cancer through mathematics: from the cell to the animal model. Nature Reviews Cancer, 10 (2010) 221--230.

\bibitem{Tindall2012} M.J. Tindall, L. Dyson, K. Smallbone, P.K. Maini. Modelling acidosis and the cell cycle in multicellular tumour spheroids. Journal of Theoretical Biology, 298 (2012) 107--115.

\bibitem{Holder2014} A.B. Holder,  M.R. Rodrigo, M.A. Herrero. A model for acid-mediated tumour growth with nonlinear acid production term. Applied Mathematics and Computation, 227 (2014) 176--198.

\bibitem{Holder2015} A.B. Holder, M.R. Rodrigo. Model for acid-mediated tumour invasion with chemotherapy intervention II: Spatially heterogeneous populations. Mathematical Biosciences, 270 (2015) 10--29.

\bibitem{Ward1997} J.P. Ward, J.R. King. Mathematical modelling of avascular tumour growth. Mathematical Medicine and Biology: A Journal of the IMA, 14 (1997) 39--69.

\bibitem{Byrne1997} H.M. Byrne, M.A.J. Chaplain. Free boundary value problems associated with the growth and development of multicellular spheroids.  European Journal of Applied Mathematics, 8 (1997) 639-658.

\bibitem{Gaffney1999} E.A. Gaffney, P.K. Maini, C.D. McCaig, M. Zhao, J.V. Forrester. Modelling corneal epithelial wound closure in the presence of physiological electric fields via a moving boundary formalism. Mathematical Medicine and Biology: A Journal of the IMA, 16 (1999) 369--393.

\bibitem{ElHachem2021} M. El-Hachem, S.W. McCue, M.J. Simpson. Invading and receding sharp-fronted travelling waves. Bulletin of Mathematical Biology, 83 (2021) 35.

\bibitem{Wiggins2003}  S. Wiggins. Introduction to Applied Nonlinear Dynamical Systems and Chaos. Second edition, Springer, New York, (2003).

\bibitem{ElHachem2021b} M. El-Hachem, S.W. McCue, M.J. Simpson. Non-vanishing sharp-fronted travelling wave solutions of the Fisher-Kolmogorov model. \href{https://arxiv.org/abs/2107.05210}{arXiv:2107.05210v2} (2021).
	
\bibitem{Du2010} Y. Du, Z. Lin. Spreading-vanishing dichotomy in the diffusive logistic model with a free boundary. SIAM Journal on Mathematical Analysis, 42 (2010) 377--405.

\bibitem{Du2011} Y. Du, Z. Guo. Spreading-vanishing dichotomy in a diffusive logistic model with a free boundary, II. Journal of Differential Equations, 250 (2011) 4336--4366.

\bibitem{Du2012} Y. Du, Z. Guo. The Stefan problem for the Fisher-KPP equation. Journal of Differential Equations, 253 (2012) 996--1035.

\bibitem{Du2014} Y. Du, H. Matsuzawa, M. Zhou. Sharp estimate of the spreading speed determined by nonlinear free boundary problems. SIAM Journal on Mathematical Analysis, 46 (2014) 375--396.

\bibitem{ElHachem2019} M. El-Hachem, S.W. McCue, W. Jin, Y. Du, M.J. Simpson. Revisiting the Fisher-Kolmogorov-Petrovsky-Piskunov equation to interpret the spreading-extinction dichotomy.  Proceedings of the Royal Society A: Mathematical, Physical and Engineering Sciences. 475 (2019) 20190378.
	
\bibitem{ElHachem2020} M. El-Hachem, S.W. McCue, M.J. Simpson. A sharp-front moving boundary model for malignant invasion. Physica D: Nonlinear Phenomena, 412 (2020) 132639.
	
\bibitem{Ward1999} J.P. Ward, J.R. King. Mathematical modelling of avascular-tumour growth II: modelling growth to saturation.  Mathematical Medicine and Biology, 16 (1999) 171--211.

\bibitem{Jin2021} W. Jin, L. Spoerri, N.K. Haass, M.J. Simpson. Mathematical model of tumour spheroid experiments with real-time cell cycle imaging. Bulletin of Mathematical Biology, 83 (2021) 44.
\cbl
\bibitem{Deutsch2005} A. Deutsch, S. Dormann. Mathematical modeling of biological pattern formation.   Birkh\"{a}user, Boston (2005).
\cb
\cbl
\bibitem{Haridas2018b} P. Haridas, A.P. Browning, J.A. McGovern, D.L.S. McElwain, M.J. Simpson. Three-dimensional experiments and individual based simulations show that cell proliferation drives melanoma nest formation in human skin tissue. BMC Systems Biology. 12 (2018) 34.\cb

\end{thebibliography}
\end{document}